\documentclass[11pt,superscriptaddress,nofootinbib]{article}
\usepackage{cite}
\usepackage{mathptmx}
\usepackage{amsmath,amsfonts,amssymb,bbm}
\usepackage{lmodern}
\usepackage{simplewick}
\usepackage[small,bf,hang]{caption}
\usepackage[bookmarks=true,hyperfigures=true,hidelinks]{hyperref}
\usepackage{slashed}
\usepackage{latexsym,epsfig}
\usepackage{color}
\usepackage{mathtools}
\usepackage{nicefrac}
\usepackage{mathrsfs}

\def\hybrid{
        \topmargin -20pt
        \oddsidemargin 0pt
        \headheight 0pt \headsep 0pt
        \textwidth 6.25in 
        \textheight 9.5in 
        \marginparwidth .875in
        \parskip 5pt plus 1pt \jot = 1.5ex}

\hybrid

 \csname
@addtoreset\endcsname{equation}{section}

\def\moth{\mathsurround=0pt}
\newdimen\zo \zo=0pt

\def\tick{\leaders\hrule height 0.5ex depth 0pt \hskip 0.5pt}
\def\upboxfill{$\moth \setbox\zo\hbox{\tick}%
  \hskip 3pt\hbox to 0pt{$\tick$\hss}\hrulefill \hbox to 7.5pt{$\tick$\hss}$}

\def\dtick{\leaders\hrule height .34pt depth 0.5ex \hskip 0.5pt}
\def\downboxfill{$\moth \setbox\zo\hbox{\dtick}%
  \hskip 2pt\hbox to 0pt{$\dtick$\hss}\hrulefill \hbox to 2pt{$\dtick$\hss}$}

\def\md{\mathrm{d}}

\def\Real{{\mathbb R}}

\def\bec{\begin{center}}
\def\ec{\end{center}}
\def\a{\alpha}  

\def\b{\beta}  
\def\c{\gamma} 
\def\C{\Gamma}
\def\d{\delta} 
\def\D{\Delta}
\def\e{\epsilon} 
\def\ve{\varepsilon}

\def\s{\sigma}

\def\t{\tau}
\def\th{\theta} 
\def\Th{\Theta}

\def\om{\omega}

\def\pa{{\partial}}
\def\Ra{{\Rightarrow}}
\def\ra{{\rightarrow}}

\def\bX{{\bf{X}}}

\def\bP{{\bf{P}}}

\def\rw{{\sqrt{w(\boldsymbol{\sigma})}}}

\def\rS{{\rm S}}

\def\gg{{\tt g}}
\def\bgg{{\bar\gg}}
\def\cL{{\cal L}}
\def\cD{{\cal D}}

\def\cO{{\cal O}}

\def\cM{{\cal M}}

\def\cH{{\cal H}}

\def\cT{{\cal T}}

\def\cO{{\cal O}}

\def\ra{\rightarrow}

\def\del{\partial}

\def\tr{{\rm tr}}
\def\Tr{{\rm Tr}\,}

 \def\det{{\rm det\,}}
\def\be{\begin{equation}}
\def\ee{\end{equation}}
\def\bea{\begin{eqnarray}}
\def\eea{\end{eqnarray}}
\def\ba{\begin{array}}
\def\ea{\end{array}}
\def\bit{\begin{itemize}}
\def\eit{\end{itemize}}

\def\mi{\mathrm{i}}
\def\me{\mathrm{e}}
\def\bs{\boldsymbol{\sigma}}
\def\bk{\boldsymbol{k}}
\newcommand{\rR}{\mathrm{R}}
\newcommand{\rT}{\mathrm{T}}
\newcommand{\bdot}{{\displaystyle\cdot}}
\newcommand{\unity}{\mathbbm{1}}

\begin{document}

\begin{titlepage}
\begin{flushright}    
{\small $\,$}
\end{flushright}
\vskip 1cm
\centerline{\huge{\bf{ A perturbative expansion scheme}}}
\vskip 0.4cm
\centerline{\huge{\bf{for supermembrane and matrix theory}}}
\vskip 1.5cm
\begin{center}
\begingroup\scshape\Large
Olaf Lechtenfeld$^{\,\diamond}$ and
Hermann Nicolai$^{\,\star}$
\endgroup
\end{center}
\centerline{{\Large ${\,}^\diamond$}\it {Institut f\"ur Theoretische Physik and Riemann Center for Geometry and Physics}}
\centerline {\it {Leibniz Universit\"at Hannover, Appelstrasse 2, 30167 Hannover, Germany}}
\vskip 0.5cm
\centerline{{\Large ${\,}^\star$}\it {Max-Planck-Institut f\"ur Gravitationsphysik (Albert-Einstein-Institut)}}
\centerline {\it {Am M\"{u}hlenberg 1, 14476 Potsdam, Germany}}
\vskip 1.5cm
\centerline{\bf {Abstract}}
\vskip .5cm
\noindent
We reconsider the supermembrane in a Minkowski background and in the light-cone gauge
as a one-dimensional gauge theory of area preserving diffeomorphisms (APDs).
Keeping the membrane tension~$T$ as an independent parameter we show
that $T$ is proportional to the square of the gauge coupling~$g$ of this gauge theory,
such that the small (large) tension limit of the supermembrane 
corresponds to the weak (strong) coupling limit of the APD gauge theory
and its SU$(N)$ matrix model approximation. A perturbative linearization 
of the supersymmetric theory suitable for a quantum mechanical path-integral treatment
can be achieved by formulating a Nicolai map for the matrix model, 
which we work out explicitly to~$\cO(g^4)$. The corresponding formul\ae\ remain
well-defined in the limit $N\ra\infty$;  this result relies on a 
cancellation of infinities not present for the bosonic membrane, indicating
that the $N\ra\infty$ limit does not exist for the purely bosonic matrix model.
Furthermore we show that the map has improved convergence properties in
comparison with the usual perturbative expansions because its Jacobian  
admits an expansion in~$g$ with a non-zero radius of convergence. Possible implications 
for unsolved issues with the matrix model of M~theory are also mentioned.
\vfill
\end{titlepage}

\section{Introduction and summary}

The maximally supersymmetric supermembrane theory in spacetime
dimension $D=11$~\cite{BST,BST1} is a model `beyond' string theory that also 
incorporates $D=11$ supergravity~\cite{CJS}, and is thus a candidate theory for a 
non-perturbative formulation of superstring theory (besides there are three
more classically consistent supermembrane theories for target-space dimensions
$D=4,5,7$ \cite{BST1}). As shown long ago~\cite{dWHN} the supermembrane in 
a flat (Minkowski) background and in the light-cone gauge can be reformulated 
as a one-dimensional maximally supersymmetric gauge theory of area-preserving 
diffeomorphisms (APDs). Building on earlier results of~\cite{G,Hoppediss, Hoppetalk} 
it has been shown that this model can be equivalently obtained as the $N\ra\infty$ limit 
of a maximally supersymmetric SU$(N)$ matrix model~\cite{dWHN}. Much later the very same model 
was re-interpreted in terms of D0 particle quantum mechanics~\cite{Witten}, and proposed 
as a model of M theory in~\cite{BFSS}. For reviews of supermembrane theory 
with many further references, see {\em e.g.}~\cite{Duff,HN,deWit}.

The main unsolved problem of (super-)membrane theory is its quantization.
Unlike for string theory there exists no gauge which linearizes the equations
of motion such that the determination of quantum correlators can be effectively
reduced to free field theory computations. Likewise, in view of the non-linearities 
a covariant path-integral approach {\em \`a la\/} Polyakov appears hopeless for either 
the bosonic or the supersymmetric membrane. A more realistically feasible
approach is based on (target-space) light-cone gauge quantization. Nevertheless, 
even with this preferred gauge choice
the solution to the problem of quantization has so far remained elusive not only 
because one has to deal with a fully interacting theory on the world volume, 
but also because it is not obvious how to set up a perturbative expansion for the 
quantized supermembrane. These difficulties are mirrored by corresponding difficulties
of the supersymmetric SU$(N)$ matrix model for $N < \infty$, as a consequence of which 
key issues remain unresolved to this day.  Apart from questions regarding the 
existence and  properties of the $N\ra\infty$ limit for the quantized theory, there are 
two main issues. One concerns the target-space Lorentz invariance of the 
quantized supermembrane (or the matrix model in the $N\ra\infty$ limit). 
For the classical theory and for finite~$N$, Lorentz invariance is in fact violated, 
but  can be recovered in the $N\ra\infty$ limit \cite{dWMN,EMM}. However, there 
has been almost no progress on the quantized theory, which would first of 
all require a proper definition of the quantized Lorentz generators and ensuring 
their quantum consistency, before actually checking the Lorentz algebra. 
Amongst other things this involves the correct definition of the light-cone
target-space coordinate~$X^-$ (which matrix theory by itself `does not know about')
as a quantum operator. Consequently, it also remains an open question whether 
$D_{\rm crit}=11$ is indeed the critical dimension for the supermembrane, eliminating the 
other classically consistent theories (see however~\cite{MKS} for some early results 
in this direction), unlike for the superstring where the well-known result $D_{\rm crit} = 10$ 
can be  established in more than one way.

A second key issue arises in connection with correlators and scattering amplitudes for 
the putative massless supermembrane excitations corresponding to the graviton, 
the gravitino and the 3-form field of $D=11$ supergravity. In particular, there is 
no (super-)membrane analog known of the Veneziano and Virasoro--Shapiro 
amplitudes, as this would almost certainly involve higher order and non-perturbative 
contributions beyond the reach of conventional string technology. Remarkably, there do exist 
classical candidate expressions for vertex operators associated to these states~\cite{DNP}, 
but like for the Lorentz generators, it has not been possible so far to turn them 
into well-defined quantum objects.  However, even if one could compute
$n$-point correlators in supermembrane theory with these vertex operators, 
there remains the more 
conceptual question as to their possible physical interpretation, and how 
such correlators would transcend the one-particle interpretation of first 
quantized  superstring theory and capture non-perturbative information.

In this paper we wish to tackle the quantization of the supermembrane
from a new and different perspective  (for a complementary 
approach to quantizing supermembrane theory, see \cite{Restuccia1}
and references therein). A main ingredient here is 
the fact that the supersymmetric APD gauge theory ({\em alias\/} the 
supersymmetric SU$(\infty)$ matrix model) {\em is\/} the supermembrane.
Our analysis leads us to the conclusion that the membrane tension~$T$
must be identified with the square of the
gauge coupling~$g$ of the APD gauge theory.
This insight allows us to set up a systematic expansion scheme 
in terms of a path-integral formulation  where the small (large) tension limit 
of the membrane theory corresponds to the weak (strong) coupling limit 
of the APD gauge theory -- a result to be contrasted with the
somewhat murky state of affairs with the zero-tension limit of string theory.
This expansion is introduced by means of a Nicolai map \cite{Nic0,Nic1}
(designated by $\cT_g$)  which we first construct for the finite-$N$ theory
up to and including quartic order $\cO(g^4)$.\footnote{
In the temporal gauge and actually including $O(g^5)$ 
since odd orders vanish at least up to this point.}
Our derivation is based on a 
systematic procedure that relies on very recent progress in perturbatively 
evaluating this map for supersymmetric Yang--Mills theories in higher dimensions, 
see especially~\cite{MN,LR2}. This prescription in principle allows for the determination
of the map to {\em any\/} desired order.  It then turns out 
that the pertinent formul\ae\ all remain well-defined in the limit $N\ra\infty$, 
via the straightforward replacement of SU($N$) commutators by APD brackets,
see especially~(\ref{TXAPD}). An accompanying $N$-dependent 
divergence cancels only for the supermembrane, indicating that the $N\ra\infty$ limit 
does not exist for the bosonic matrix model~\footnote{
Related difficulties with the $N\ra\infty$ limit for the bosonic matrix model 
were already pointed out in \cite{Sav}.}
(`non-renormalizability' of the bosonic
membrane). In arriving at this conclusion, the map $\cT_g$ plays a crucial role:
as far as we are aware, there is no other approach that addresses the issue
of the $N\ra\infty$ limit in such a direct manner. These are two main results 
of this paper which should eventually permit setting up an approximation scheme also for 
correlators and other quantities of physical interest. This might enable one to sidestep 
the finite-$N$ approximation altogether and to deal directly with the limiting theory
for $N=\infty$. 

Finally, we demonstrate that the expansion of the Jacobian in 
powers of~$g$ has a {\em non-zero\/} radius of convergence. This does not
yet mean that the map itself has this property, but it represents strong evidence 
that the perturbative expansion for $\cT_g$ is indeed better behaved 
than the usual quantum field theoretic perturbation expansions. 
Of course, employing this map does not alter the standard perturbative expansion
of quantum correlation functions. However, it breaks it up into two stages:
In the first step, the field arguments in the correlator are replaced by the ones
transformed with the inverse map $\cT_g^{-1}$, 
which is given in terms of a tree-graph expansion. In the second step, 
one performs a free-field correlator of those tree graphs, which finally 
produces loop diagrams. Therefore, the perspective is that the tree 
expansion of the first stage is  better than asymptotic (even convergent),
while the divergent field-theoretic high-order perturbative behavior 
gets encoded in the free-field correlators of the second stage (see \cite{NP} 
for a sample calculation along these lines for $D=4$ maximal super-Yang--Mills
theory). Let us also note the existence of a closed-form expression for $\cT_g$ 
in terms of a path-ordered integral, that has no analog in standard 
perturbative quantum field theory~\cite{LR1}. In particular, this formula directly 
reproduces the polynomial form of $\cT_g$ in those cases where such a form 
is known to exist (such as for supersymmetric quantum mechanics), so
its further exploration might well reveal unknown special
properties of maximally supersymmetric matrix theory as well.

To be sure, there still remains a long way towards a full quantum
treatment of the supermembrane, both on the technical and on the conceptual
side. Nevertheless, the present results offer novel perspectives on this old problem
and open several new avenues for future research.
A particularly intriguing one concerns  the fact that the map $\cT_g$ transforms 
the interacting functional  measure of the APD gauge theory
to a free one which is ultralocal in the membrane coordinates.
Although this complicates setting up a sensible perturbative expansion for 
small membrane tension, there is a remarkable analogy here with the small-tension 
limit of M theory considered in~\cite{DHN}, where it is the BKL limit which 
likewise leads to an ultralocal theory via the causal decoupling of spatial points.
The latter limit has been shown to
exhibit hints of the maximal-rank hyperbolic Kac--Moody algebra E$_{10}$ 
as a fundamental symmetry.  It is a fascinating challenge to explore the 
possible links between these {\em a priori} different versions of the small-tension 
limit, and thus to reconcile two very different perspectives on M theory.

The structure of this paper is as follows. In Section~2 we review basic results 
on the light-cone gauge formulation of the supermembrane, mostly following
the exposition in~\cite{dWHN}, and set up the path integral in Subsection~2.3.
Section~3 explains the construction of the Nicolai map both for the matrix theory and 
the APD gauge theory. This section contains our main result, namely an explicit
form of the map~$\cT_g$ in an expansion to quartic order in the coupling~$g$. 
The question of the behavior of the map in the complex $g$~plane (and thus 
the convergence properties of the expansion) is addressed in Section~4.
Finally an Appendix provides details of two consistency checks on the main 
result derived in this paper.

\section{Supermembrane basics}

In the section we closely follow~\cite{dWHN} to which we refer for further details of
the derivation. The main difference is that we here keep the membrane tension~$T$ 
as an independent parameter, in order to expose the link with weakly and strongly 
coupled Yang--Mills theory.

\subsection{Supermembrane in the light-cone gauge with variable tension}

Classically consistent supermembranes exist for target-space dimensions
$D=4,5,7$ and 11, but in the remainder we will restrict attention
mostly to the maximally supersymmetric case, for which $D=11$. 
The target superspace coordinates 
$\big\{X^\mu, \th\big\}  \equiv \big\{ X^\mu (\xi^i)\,,\, \th(\xi^i) \big\}$
with the range $\mu,\nu,\ldots=0,1,\ldots,D-1 = 10$ 
are then functions of the membrane  world-volume coordinates 
\be
(\xi^i) \equiv (\t,\bs) \equiv (\t,\s^r)  
\qquad\textrm{where}\quad i,j,\ldots=0,1,2 \quad\textrm{and}\quad r,s,\ldots=1,2\ .
\ee
$\th(\t,\bs)$ is a real 32-component Majorana spinor of SO(1,10) (we usually do
not write out spinor indices). The target-space vielbein is
\be
E_i{}^\mu\ =\ \del_i X^\mu + \bar \th\,\C^\mu \del_i \th \ .
\ee
For $D=11$, the real 32-by-32 $\C$-matrices generate the SO(1,10) Clifford algebra, 
$\{\C^\mu , \C^\nu \} = 2\eta^{\mu\nu}$. The world-volume metric is
\be
\gg_{ij}\ \equiv\ E_i{}^\mu E_j{}^\nu \eta_{\mu\nu}\ .
\ee
For the light-cone gauge we split the target-space coordinates as
\be
\{ X^\mu \} \equiv \{ X^+,X^-,X^a \}  \qquad\quad \textrm{with}\qquad
X^\pm = \tfrac1{\sqrt{2}} (X^{10} \pm X^0) \qquad\textrm{and}\qquad
\{ X^a\} \equiv \bX
\ee
being the transverse components $( a,b,\ldots=1,\ldots,9)$.
We adopt the target-space light-cone gauge
\be\label{LCG}
X^+(\t,\bs) \ =\  X_0^+ \,+\, \t \ , \qquad \C_+ \th(\t,\bs)\ =\ 0
\ee
thus identifying the target-space light-cone coordinate $X^+$ with the world-volume time
coordinate $\t$. With these gauge choices the induced metric on the three-dimensional world
volume is 
\be\label{3metric}
\begin{aligned}
\gg_{rs} &\ \equiv\ \bgg_{rs}\ =\ \del_r \bX \cdot \del_s \bX \ ,\\
\gg_{0r} &\ \equiv\ u_r\ =\ \del_r X^- + \del_0 \bX\cdot\del_r \bX + \bar\th\,\C_- \del_r \th \ ,\\
\gg_{00} &\ =\ 2\del_0 X^- + (\del_0\bX)^2 + 2\bar\th\,\C_- \del_0 \th\ .
\end{aligned}
\ee
The metric determinant is 
\be
\gg \ \equiv\, \det \gg_{ij} \ =\  - \D\,\bgg
\ee
with 
\be
\bgg\ \equiv\ \det \bgg_{rs} \qquad\textrm{and}\qquad
\D\ =\ - \gg_{00} + u_r \bgg^{rs} u_s  \ ,\qquad \bgg^{rs} \bgg_{st} = \d^r_t\ .
\ee
The supermembrane Lagrangian then becomes
\be\label{Lag}
\cL \ =\  T\,\Bigl(- \sqrt{\bgg\,\D} \,+\, \e^{rs} \del_r X^a \bar\th\,\C_- \C_a \del_s \th\Bigr)\ ,
\ee
where we now include the membrane tension $T$ as an independent parameter.
In principle the membrane tension is of dimension [mass]$^3$, but we
here find it convenient to render all variables dimensionless by rescaling them 
with appropriate powers of some reference mass scale (as was already 
implicitly assumed in (\ref{LCG})). This reference scale has no physical 
meaning in and by  itself, as a proper identification of the gravitational 
coupling (Newton constant or Planck mass) and evaluating its relation to $T$
will require the evaluation of a graviton scattering amplitude, 
as is the case in string theory. However, for the doubly 
dimensionally reduced supermembrane~\cite{Duff1} such a 
relation can indeed be established by noting that $T R_{10} = T_s
\equiv (2\pi\alpha')^{-1}$, where $T_s$ is the string tension and $R_{10}$ 
the radius of the compactified 11th dimension. Because the latter is related 
to the string coupling by $R_{10}^3 T = g_s^2$~\cite{Witten1,T},
we see that $T^2=T_s^3g_s^{-2}$, or
\be\label{Tgs}
T\ =\ g_s^{-1}\,(2\pi\a')^{-3/2} \ .
\ee
In this way the parameter $T$ ties together the two key parameters of string theory, 
and thus also with the APD gauge coupling $g$  via (\ref{T=g}) below. This will
eventually allow us to study how the standard perturbative expansions in string
theory are related to the expansion in the APD Yang-Mills coupling that we
will derive in section 3.3.

With these conventions the (dimensionless) canonical momenta are
\be\label{momenta}
\begin{aligned}
P^+ &\ =\ T \sqrt{\frac{\bgg}{\D}}\ ,    \\
\bP &\ =\ \frac{\d\cL}{\d \del_0\bX}\ =\ T \sqrt{\frac{\bgg}{\D}}
\big( \del_0 \bX - u_r \gg^{rs} \del_s \bX \big) \ \equiv\
P^+ \big( \del_0 \bX - u_r \gg^{rs} \del_s \bX\big)\ ,  \\
S &\ =\  \frac{\d\cL}{\d \del_0\bar\th} \ =\
- T \sqrt{\frac{\bgg}{\D}}  \C_- \th\ \equiv\ - P^+ \C_- \th\ .
\end{aligned}
\ee
The last formula implies a second-class constraint (entailing the replacement of Poisson 
brackets by Dirac brackets). The formul\ae~(\ref{momenta}) imply the first-class constraint
\be
\phi_r \ =\  \bP\cdot \del_r \bX \,+\, P^+ \del_r X^- \,+\, \bar{S} \del_r \th \ \approx\ 0\ ,
\ee
which generates spatial diffeomorphisms on the membrane. This gauge freedom can
be exploited to set $u^r = 0$ in (\ref{3metric}), which in turn implies
\be
\del_r X^- \ =\ - \del_0 \bX\cdot\del_r \bX \,-\, \bar\th\,\C_- \del_r \th  \ .
\ee
To be able to solve this equation for $X^-$ we must impose the integrability constraint
\be\label{APD0}
\phi \ \equiv \  \e^{rs} \big( \del_r \del_0 \bX\cdot \del_s \bX \,+\, \del_r \bar\th\,\C_- \del_s\th\big)
\ \approx \  0\ .
\ee
This constraint generates APDs on the membrane: while general (spatial) diffeomorphisms
on the membrane are generated by vector fields $\d\xi^r(\bs)$, APDs are generated by 
divergence-free vector fields obeying $\del_r (\sqrt{w}\d\xi^r) = 0$ (where the reference
density $\rw$ coincides with the one introduced in (\ref{P+}) below). The latter are 
locally of the form $\sqrt{w} \d\xi^r = \e^{rs}\del_s \d\xi$ with a scalar 
parameter $\d\xi(\bs)$. On higher-genus 
membranes there are in addition topologically non-trivial diffeomeophisms formally generated 
by harmonic vector fields~\cite{dWMN}, which we will, however, disregard here.

With these gauge choices the (dimensionless) Hamiltonian density becomes (see also~\cite{Tanii})
\be\label{M2}
\begin{aligned}
\cH(\bs) \ &\equiv \  - P^-(\bs) \ =\ \bP\cdot \del_0 \bX \,+\, P^+ \del_0 X^-\, +\, \bar{S}\del_0 \th 
\,-\, \cL \\
&=\ \frac{\bP^2 \,+\, T^2 \bgg}{2P^+} \ -\  T\e^{rs} \del_r X^a \bar\th\,\C_- \C_a \del_s \th
\end{aligned}
\ee
whose bosonic part was already derived long ago in~\cite{G,Hoppediss, Hoppetalk} (for $T=1$).
Here we see why we must choose the membrane tension  to be positive; 
flipping the sign of~$T$ will change 
the sign of the kinetic part of the Hamiltonian by~(\ref{momenta}), 
hence result in an instability. This is, of course, in accord with expectations.

Because $P^+(\t,\bs)$ obeys the Hamiltonian equation of motion $\pa_\t P^+ (\t,\bs) = 0$
and transforms as a density we can set~\cite{dWHN}
\be\label{P+}
P^+(\t, \bs)\ =\  P^+_0 \rw
\ee
where $P^+_0>0 $ is constant, and $\rw >0$ is a reference density 
normalized to $\int\!\md^2 \s \, \rw = 1$ (with an associated
reference metric $w_{rs}(\bs)$ on the membrane, which is however only needed when
discussing target-space Lorentz invariance~\cite{dWMN}).
This leads to the (dimensionless) mass operator
\be\label{M1}
\cM^2 \ =\  - 2 P^+_0 P^-_0  - \bP^2_0 \ = \ 
\int\!\md^2\s \ \left( [\bP^2]' \,+\, T^2 \bgg  \,-\,
2T\e^{rs} \del_r X^a \bar\Th\,\C_- \C_a \del_s \Th \right) 
\ee
with $P^-_0 = \int\!\md^2 \s\,P^-(\bs)$ and rescaled fermionic variables~\footnote{
Which obey the canonical (Dirac)  brackets 
$\{ \Theta(\bs) \,,\, \bar\Th(\bs')\}_{\rm DB} = (4\rw)^{-1} \C_+ \d^{(2)}(\bs{,}\bs')$ \cite{dWHN}.}
\be
\Th (\bs)\ \equiv \  \sqrt{P^+_0}  \, \th(\bs)\ .
\ee
The prime in (\ref{M1}) indicates that zero modes have been removed 
from $\int\!\md^2\s \, \bP^2(\bs)$. 

Because the zero modes $X^a_0$ and $\th_0$ decouple from the 
Lagrangian (\ref{Lag}), the eigenstates of the mass operator take the form of a 
direct product of the  massless $D=11$ multiplet ($\bf{44} \oplus {\bf 84}$ bosonic 
and $\bf{128}$ fermionic states) with an eigenstate of the mass operator (\ref{M1}) 
\cite{dWHN}. For the uncompactified supermembrane the spectrum 
of the latter is known to be continuous \cite{dWNL,Smilga} (but becomes discrete 
after compactification with winding \cite{Restuccia2}). This fact can be interpreted
as evidence that the supermembrane does not admit a first-quantized formulation,
but must be regarded as a non-perturbative theory from the outset \cite{HN}.

Finally we note that the fulfilment of the constraint (\ref{APD0}) allows us
to solve for the target-space coordinate $X^-$: we have
\be
X^-(\t, \bs) \ =\ - \int\!\md^2\s' \ 
G^r(\bs,\bs') \Big( \del_0 \bX\cdot\del_r \bX(\t,\bs')  \,-\, \bar\th\,\C_- \del_r \th(\t,\bs')\Big)
\ee
with a suitable Green's function obeying $\pa_r G^r(\bs,\bs') = \d(\bs{,}\bs')$
\cite{G,dWMN}. This formula is needed for the target-space boost 
generators and for the verification of target-space Lorentz invariance
in the classical limit~\cite{dWMN,EMM}. It is worth  pointing out that this
information is {\em not} available in the matrix model as such, where the Lorentz 
boost generators must either be ``guessed" or deduced from the supermembrane 
matrix-model correspondence, as in~\cite{dWMN}.

\subsection{APD gauge theory and matrix model}

With the above formula for the mass operator the supermembrane theory can be 
reformulated as a one-dimensional supersymmetric gauge theory of area preserving diffeomorphisms~\cite{dWHN}.
This can be seen by exploiting the algebraic identity
\be
\bgg \ =\  \det \big( \del_r\bX\cdot \del_s\bX \big)\ =\ 
  \big\{ X^a, X^b\big\}  \big\{ X^a, X^b\big\}\ ,
\ee
where the APD bracket of any two functions $A(\bs)$ and $B(\bs)$ on the membrane is defined by
\be\label{APDbracket}
\big\{ A(\bs)\, , \, B(\bs) \big\} \ :=\ \frac1{\rw}\,\e^{rs} \del_r A(\bs)  \del_s B(\bs)\ .
\ee
This is indeed a Lie bracket (obeying antisymmetry and the Jacobi identity)~\cite{G,Hoppediss, Hoppetalk}.

Then (\ref{M2}) can be equivalently obtained from the supersymmetric 
Lagrangian~\footnote{
While $\t$ is the time coordinate on the membrane world-volume,
we denote the Yang--Mills time coordinate by $t$, but keep the erstwhile membrane
coordinates $\s^r$ as labels for the APD gauge group.}
\be\label{LYM}
\tfrac1{\sqrt{w}} \,\cL \ =\  
\tfrac12 (D_t \bX)^2 \ +\ \bar\Th\,\C_- D_t\Th \ -\ 
\tfrac14 g^2 \{ X^a,X^b\}^2\ +\  g\,\bar\Th\,\C_- \C_a \{X^a , \Th\} \ .
\ee
The comparison between $T$ and the Yang-Mills coupling $g$ is a bit more
subtle. Putting dimensions back in we see that in one (time) dimension
the Yang-Mills coupling $g$ has dimension [mass]$^{3/2}$, whence we must
identify
\be\label{T=g}
T \; \propto \;  g^2 \ .
\ee
In view of our comments after (\ref{M2}) we see that negative $T$ would
make $g$ imaginary, again indicating that $T>0$ is required for
stability of the membrane.
Hence the small (large) tension limit of the supermembrane corresponds
to the weak (strong) coupling limit of the supersymmetric APD gauge theory.
The APD covariant derivative is given by
\be
D_t f (t,\bs) \ := \ \del_t f (t,\bs)  \,+\, g\,\big\{ \om (t,\bs) \, ,\,  f(t,\bs) \big\}
\ee
with the APD gauge field $\om(t,\bs)$ which is here introduced {\em ad hoc}, as it is 
absent from the supermembrane action. The Lagrangian (\ref{LYM}) is nothing but the 
dimensional reduction of maximally extended super-Yang--Mills theory~\cite{SYM1} to one (time) 
dimension, with the identifications $\om \equiv A_0$ and $X_a \equiv A_a$, and $g$ 
the usual Yang--Mills  coupling, but now with the infinite-dimensional APD gauge group.
This works precisely in the dimensions where pure supersymmetric Yang--Mills theories 
exist, namely $D=3,4,6,10$ \cite{SYM1}, in agreement with the admissible target-space
dimensions 4,5,7 and 11 for supermembranes.

The group of (homotopically trivial)
APDs on the membrane can be approximated by the finite-dimensional 
unitary groups SU($N$), such that the full group is recovered in the limit 
$N\ra\infty$ \cite{G,Hoppediss, Hoppetalk}. Replacing APDs by SU($N$) gives the matrix model of M theory.
For this approximation one expands all functions on the membrane into a complete 
orthonormal set of functions $Y^A(\bs)$, 
\be
 \int\!\md^2\s \ \rw\,Y^A(\bs)\,Y^B(\bs) \ =\ \d^{AB} \ ,
\ee
where we separate off the zero modes,
\be
\begin{aligned}\label{ExpandAPD}
X_a(t,\bs) \ &=\  X_a^{(0)}(t) \ +\  \sum_{A=1}^\infty X_a^A(t) Y^A(\bs) \ ,\\
\om(t,\bs) \ &=\  \om^{(0)}(t) \ +\  \sum_{A=1}^\infty \om^A(t) Y^A(\bs) \ ,\quad \\
\Th(t,\bs) \ &=\  \Th^{(0)}(t) \ +\  \sum_{A=1}^\infty \Th^A(t) Y^A(\bs) \ .
\end{aligned}
\ee
The zero modes $X_a^{(0)}(t)$ and $\Th^{(0)}(t)$ decouple in (\ref{LYM}), where
$X_a^{(0)}(t)$ describes the center of mass motion of the membrane as a whole.  
Likewise, the gauge zero mode $\om^{(0)}(t)$ drops out in the Lagrangian (as it acts
effectively like a  U(1) gauge field, which cannot couple because both 
$X_a^{(0)}$ and $\Th^{(0)}$ are real). The remaining non-zero modes
describe the `internal' degrees of freedom of the supermembrane. The APD gauge group
can thus be approximated by  SU$(N)$, as is most easily and explicitly 
done for $S^2$ \cite{G,Hoppediss, Hoppetalk} and  $T^2$ \cite{Hoppe1,FFZ,dWMN}, by cutting off 
the mode expansions at $N^2{-}1$ (ignoring topological modes) and replacing
the APD-brackets by SU($N$) commutators. In fact, as shown in~\cite{BMS} the SU($N$) 
approximation works for {\em any} genus of the membrane.  Consequently, we have
\be\label{fABC}
f^{ABC}_{\rm APD} \ \equiv \  
\int d^2\s\, \rw \, Y^A(\bs) \big\{ Y^B(\bs) , Y^C(\bs)\big\}\ =\
\lim_{N\ra\infty} f^{ABC}(N)
\ee
with SU($N$) structure constants $f^{ABC}(N)$. Hence the expansion labels
$A,B,\ldots=1,\ldots,N^2{-}1$ are thus turned into Yang--Mills indices, while $a,b,\ldots$ are 
transverse (for membrane) and space-like (for supersymmetric Yang--Mills) indices.

After these preparations,  the matrix-model Lagrangian assumes the standard 
form~\footnote{
For finite-dimensional gauge groups these supersymmetric matrix
models were first obtained in~\cite{CH,BRR,Flume}.}
\be\label{MMLag}
\cL \ =\  \tfrac12 (D_t X^A_a)^2 \,-\, \mi\,\th^A_\a D_t \th^A_\a \,- \, 
 \tfrac14 g^2 (f^{ABC} X^B_b X^C_c)^2 \,-\, 
 \tfrac{\mi}{2}\,g\,f^{ABC} \th^A_\a \c^a_{\a\b} X^B_a \th^C_\b\ ,
\ee
where we have now switched to SO(9) spinors $\th^A_\a$ with 16 real components 
and where
\be
D_t \th^a\ =\ \del_t \th^A + g\, f^{ABC} \om^B \th^C 
\ee
is the SU($N$) covariant derivative. The real symmetric 16-by-16 matrices 
$\c_a$ generate the SO(9) Clifford algebra, $\{\c^a , \c^b \} = 2\d^{ab}$.
Variation w.r.t.~$\om^A$ yields the constraint
\be
f^{ABC} \big( X^B_a D_t X^C_a + \th^B_\a \th^C_\a \big) \ \approx \  0
\ee
which is equivalent to the canonical generator of SU$(N)$ gauge transformations
(after performing this variation we can put $\om^A =0$ everywhere). 
The Lagrangian (\ref{MMLag}) is the one that underlies the M theory conjecture
of \cite{BFSS}, see also \cite{Bigatti,Banks,Taylor}.

Of course, the group of area-preserving diffeomeorphisms
also depends on the topology of the membrane. For topologically
non-trivial membranes, the APDs continuously connected to the identity 
constitute a normal subgroup APD$_0$ within the 
group of all area-preserving diffeomorphisms \cite{dWMN}.
It is only the subgroup APD$_0$ that can be approximated 
by SU($N$) (indeed for arbitrary genus of the membrane \cite{BMS}), 
whereas diffeomorphisms in the quotient  APD$/$APD$_0$ are beyond 
the reach of the SU($N$) matrix approximation.\footnote{
For fixed genus, the quotient APD$/$APD$_0$ is nothing but the mapping class 
group of the membrane (we are grateful to A. Kleinschmidt for a 
discussion on this point).}
The fact that one can interpolate 
between membranes of different topology by means of thin tubes that cost no energy
(see {\em e.g.} \cite{HN} for an explanation of this point) then raises 
the question of how to accommodate different topologies in a single 
unified APD formulation, and suggests adopting a `universal' APD group 
encompassing membranes of all genera, perhaps along the lines of \cite{FS}.

\subsection{Setting up the path integral}

Our goal is now to set up a path-integral formulation that should eventually 
permit the computation of correlators of physically  relevant quantities, and
complement the canonical quantization methods underlying many treatments
of the matrix model. We shall thus be interested in evaluating correlation functions
of the type
\be\label{Corr1}
\begin{aligned}
\Big\langle \cO_1 \cdots \cO_n \Big\rangle_g  \ &=\
\int\!\prod  \cD X_a(t,\bs)\,\cD\th_\a(t,\bs)\,\cD\om(t,\bs)\,\cD C(t,\bs)\,\cD\bar{C}(t,\bs) \ \times \\
&\qquad\qquad\times \ \cO_1[\bX,\th] \cdots \cO_n[\bX,\th]\ \exp\bigl( \mi\,\rS_{\rm tot} \bigr)
\end{aligned}
\ee
where the precise form of the functionals $\cO_i[\bX,\th]$ need not be specified at
this point. Because this is a gauge theory, the full action
\be
\rS_{\rm tot}\ =\  \rS \,+\, \rS'
\ee
with $\rS = \int\!\md t\,\cL$  must comprise a gauge-fixing part $\rS' = \int\!\md t\,\cL'$.
For higher-dimensional Yang--Mills theories there are two preferred choices,
namely the Lorenz gauge $\pa^\mu A_\mu = 0$, and the axial gauge
$n^\mu A_\mu = 0$ (which includes the light-cone gauge for null vectors $n^\mu$).
In the reduction to one time dimension the axial gauge is necessarily identical
with the temporal gauge. Consequently, we have two preferred choices for the 
gauge-fixing part, namely
\be
\begin{aligned}
\cL'\ &= \ -\tfrac{1}{2\xi}(\del_t\om)^2\,+\,\bar{C}\,\del_t D_t C
\quad\qquad\, \textrm{(Lorenz gauge)} \ ,\\[4pt]
\cL'\ &= \ -\tfrac{1}{2\xi}\om^2\,+\,\bar{C} D_t C
\quad\qquad\qquad  \textrm{(temporal gauge)} \ .
\end{aligned}
\ee
A further peculiarity of one dimension is that the temporal gauge implies the Lorenz gauge
\be\label{axialLorenz}
\om(t,\bs)  = 0 \qquad \Ra \qquad \pa_t \om(t,\bs) \equiv \dot{\om}(t,\bs) = 0 \ .
\ee
$C(t,\bs)$ and $\bar{C}(t,\bs)$ are the usual Faddeev--Popov ghosts~\cite{FP,tH},
and $\xi$ is a real parameter which will be eventually sent to zero to put the theory
on the gauge hypersurface.  After trading the $\bs$~dependence for SU$(N)$ indices,
we are left with a quantum mechanical path integral describing {\em finitely\/} many
degrees of freedom. Because of the supersymmetry there is no need for a 
normalization factor in (\ref{Corr1}) (as can be easily checked for $g=0$ with both 
gauge choices). In passing we note that we can of course equivalently switch to a 
Euclidean formulation by flipping  the sign in the kinetic terms $(\dot X_a)^2$, $\dot\om^2$ 
and for the ghosts, and by replacing the oscillatory exponent by $\exp(-\rS_{\rm tot})$; 
the factor $\mi$ is then absent in the spinor kinetic term.

An important part of our construction is that we consider the path
integral in a form where the fermions (and also the ghosts) are integrated out.  
For the temporal gauge and for finite~$N$ the integration over $\th^A_\a(t)$ results in the 
Matthews--Salam--Seiler~(MSS) determinant~\cite{MS,S}
\be\label{MSS}
\D_{\rm MSS}\big[\om{=}0, \bX\big]  \ =\  \Bigl[
\det \bigl( \d^{AB} \d_{\a\b}\,\d(t_1{-}t_2)\ +\ g\,K^{AB}_{\a\b}(t_1,t_2)\bigr) \Bigr]^{1/2} 
\ee
which is actually a Pfaffian because we are integrating over {\em real\/} fermions.
The integral kernel appearing in this expression is
\be \label{Kkernel}
K^{AB}_{\a\b}(t_1,t_2)\ :=\ 
\ve(t_1{-}t_2)\,f^{ACB}\c^a_{\a\b} X_a^C(t_2) \ .
\ee
This is a real operator which is however not symmetric because hermitian 
conjugation also exchanges the arguments $t_1$ and $t_2$.
Furthermore, we have taken out trivial factors of $\det (\pa_t)$ (which
anyway cancel in the supersymmetric path integral).
The free fermion propagator~$\ve$ is just the Green's function for~$\del_t$,
\be \label{freeprop}
\ve(t-t')\ =\ \bigl[\del_t^{-1}\bigr](t,t') \ =\
\int\!\frac{\md p}{2\pi}\ \frac{\mi p}{p^2 - \mi\e} \me^{-\mi p(t-t')} \ =\
\Theta(t{-}t')-\tfrac12 \ =\ -\ve(t'{-}t)\ .
\ee
This choice of integration constant implies $\ve(0) = 0$ as well as
\be
\int\!\md t \  \ve(t{-}t') \ =\  0\ .
\ee
Our particular choice is important for the tests in the Appendix which 
otherwise cannot be satisfied (it is also consistent with the dimensional 
reduction of the usual Dirac propagator). In Section~4 we will study some properties 
of this determinant in more detail and prove in particular that the expansion of 
$\log(\D_{\rm MSS})$ in powers of~$g$ has a {\em non-zero radius of convergence\/} 
with suitable technical assumptions on the behavior of~$X_a^A(t)$. We also note that 
we have no positivity statement about~$\D_{\rm MSS}$ (though the fermion determinant is 
non-negative for complex fermions!). Similarly, the determinant cannot be shown to be 
an even function of~$g$ because of the non-vanishing trace 
${\rm tr}\,(\c^{a_1}\cdots\c^{a_9})=16\,\e^{a_1\cdots a_9}$.

For the infinite-dimensional APD gauge group we must be a little more careful:
while the kinetic term of (\ref{LYM}) is local in $\bs$, the interaction term is not 
because it contains derivatives in $\bs$. To take into account this non-locality 
we can formally replace the integral kernel (\ref{Kkernel}) by
\be \label{KkernelAPD}
K^{\rm APD}_{\a\b}(t_1,t_2\,;\,\bs_1,\bs_2)\ :=\ 
\ve(t_1{-}t_2)\,\c^a_{\a\b} \, \frac{1}{\sqrt{w(\boldsymbol{\sigma_1})}}\,
\e^{rs}\,\frac{\pa X_a(t_2,\bs_1)}{\pa \s^r_1}\,
\d(\bs_1,\bs_2)\,\frac{\pa}{\pa \s^s_2}   
\ee
and the identity operator by $\d_{\a\b}\d(t_1{-}t_2)\,\d(\bs_1,\bs_2)$, with the
proviso that folding with this kernel now also contains an integral over~$\bs$.
When expanding the logarithm of the MSS~determinant using log det = tr log,
we encounter for each trace a divergent factor $\d(\bs,\bs)$. 
This factor corresponds to a factor of~$N$ arising in the matrix-model regularization 
for each trace over the Yang--Mills indices (as in $f^{ACD} f^{BCD} = N\,\d^{AB}$),
which also diverges in the limit $N\ra\infty$.\footnote{
To get the proper APD structure constants in the $N\ra\infty$ limit, one must adopt a 
suitable normalization of the SU($N$) generators, see {\em e.g.\/} appendix A of~\cite{dWMN}.}
Importantly, for the supersymmetric theory
this divergence is compensated by a corresponding factor from the Jacobian,
as follows directly from (\ref{detrelation}). 
This cancellation explains why our final result~(\ref{TXAPD}) 
is perfectly well-defined. At the same time it indicates that 
the $N\ra\infty$ limit does not exist for the purely bosonic matrix model, thus giving
meaning to the statement that the `bosonic membrane is non-renormalizable'.

\subsection{Physical correlators}

With the path integral formalism at hand we can now in principle proceed to calculate gauge-variant
and gauge-invariant correlators of suitable objects. But what are the physically 
relevant operators $\cO[\bX,\th]\,$? As in string theory, for the membrane the latter 
should be associated to vertex operators describing the emission or absorption of certain
one-particle excitations from the membrane. As first shown in~\cite{DNP} there 
indeed exist the classical analogs of supermembrane light-cone vertex operators
exciting the massless states of the supermembrane, which comprise 
the massless supermultiplet of maximal $D=11$ supergravity~\cite{CJS}. 
The related expressions must satisfy various consistency 
constraints (target-space and world-volume gauge invariance, linear and non-linear 
supersymmetry) which are explained at length in~\cite{DNP}, corresponding to (but
more complicated than) the ones known from type~II superstring theory. In particular,
in analogy with closed-string vertex operators they are to be integrated over 
the membrane world volume.
For instance, for the transverse graviton components we have \cite{DNP}
\be
\cO[\bX,\th] \ =\ \int\!\md t\,\md \bs \ V_h[\bX,\th]
\ee
with
\be
\begin{aligned}
V_h[\bX,\th] \ &=\ h_{ab} \Bigl[ D_tX^a D_t X^b\ -\ \{X^a,X^c\}\{X^b, X_c\}
\ -\ \mi\bar\th\,\c^a \{ X^b , \th\} \\
&\quad
-\ \tfrac12 D_t X^a \,\bar\th \c^{bc}\th \, k_c\ -\ \tfrac12 \{X^a , X_c\} \, \bar\th \c^{bcd}\th \, k_d
\ +\ \tfrac18 \bar\th\c^{ac}\th \, \bar\th \c^{bd}\th \, k_c k_d \Bigr]
\ \me^{-\mi\bk\cdot\bX+\mi k^- t} 
\end{aligned}
\ee
where $h_{ab}$ is the transverse graviton polarization tensor, 
and $\{k_a\}=\bk$ denotes the transverse components of the target-space momentum.
For the light-cone gauge target-space 
momentum $k^\mu$ one must furthermore assume $k^+ = 0$ in order to avoid having to 
deal with the light-cone coordinate $X^-(\t,\bs)$ in the exponential (as is also customary 
in string theory~\cite{GS}). Remarkably, and unlike for superstring theory, there do not 
appear to exist analogs of the string vertex operators for massive string states.
This would be in accord with the fact that the supermembrane is not a first quantizable 
({\em i.e.}~one-particle) theory~\cite{HN} and for finite~$N$ consistent with the 
D0-multiparticle interpretation of~\cite{BFSS}.

Because the light-cone vertex operators are given by complicated expressions,
and because the measure in (\ref{Corr1}) is not Gaussian, no sustained attempt has been 
made, as far as we are aware, to evaluate their correlators. Neither has it been
possible so far to set up a perturbative expansion, as this will also require 
understanding the quantum corrections (renormalizations) that are necessary for 
the vertex operators to remain well-defined in the quantized interacting theory. Our line 
of attack will therefore be a different one, in that we will reformulate the above path integral 
in terms of a Nicolai map. A main advantage of such an approach is that the formul\ae\
to be presented below remain perfectly well-defined in the  limit $N\ra\infty$ and can 
thus be consistently implemented also in the APD path integral. Consequently,  it may be possible 
in this way to sidestep the  detour via the finite-$N$ matrix model, and to directly 
tackle the $N=\infty$ theory  right away. Possible applications of this technology 
to supermembrane vertices will, however, be left to future work.

\section{The map to fourth order}

The method that we propose here to tackle expressions like (\ref{Corr1}) is based on the 
Nicolai map $\cT_g$ \cite{Nic0,Nic1,Nic2,FL,DL1,DL2,L1}, exploiting recent progress 
in determining this map to higher orders in $g$ \cite{ANPP,ALMNPP,MN,LR2}.
This map is a non-local and non-linear field transformation, which
maps the theory to a free theory in such a way that after integrating out the fermions 
(gaugini and ghosts) the product of the resulting fermionic determinants 
equals the Jacobian of the map $\cT_g$ at least locally in field space.
For operators $\cO_k(t_k)$ built from $X_a$ (and~$\om$) alone, 
this enables us to re-express the expectation value (\ref{Corr1}) in the matrix theory as 
a free-field correlator of transformed bosonic fields, {\em viz.}
\be\label{Corr2}
\Big\langle \cO_1(t_1) \cdots \cO_n(t_n) \Big\rangle_g \ =\ 
\Big\langle \cT_g^{-1}(\cO_1(t_1)) \cdots \cT_g^{-1}(\cO_n(t_n)) \Big\rangle_0 
\ee
where integrating out the gaugini and ghosts is trivial on the right-hand side
because the transformed operators are purely bosonic ones.
We can therefore read this relation as one in the integrated-out theory 
as well as in the original one including the fermions.\footnote{
An extension to include fermionic or ghost arguments in $\cO_k$
is straightforward but renders them nonlocal in the integrated-out theory.}
A key property
of the map $\cT_g$ is the equality of its functional Jacobian with the
product of the fermionic determinants obtained by integrating out
all anticommuting variables, to wit,
\be \label{detrelation}
\det \left( \frac{\d \cT_g X}{\d X} \right) \ =\
\D_{\rm MSS}[\om,\bX]\ \D_{\rm FP} [\om,\bX]
\ee
where $\D_{\rm FP}$ and $\D_{\rm MSS}$ are, respectively, 
the Faddeev--Popov determinant~\cite{FP,tH} and the 
MSS~determinant~(\ref{MSS})~\cite{MS,S}.
We refer readers to \cite{ANPP,ALMNPP,LR1,MN,LR2}
for recent progress in constructing the map $\cT_g$ for pure supersymmetric
Yang--Mills theories in all relevant dimensions. A crucial simplification follows from
(\ref{axialLorenz}), since it allows us to largely ignore the distinction between 
`on-shell' and `off-shell' $R$-prescriptions in \cite{MN,LR2} that must be taken into 
account in more than one dimension.

Computing quantum correlation functions via (\ref{Corr2})  may in particular shed 
new light onto the combinatorial divergences appearing in higher orders 
of perturbation theory, due to the separation of the computation into two stages.
The first step amounts to writing out the operators appearing on the right-hand side
of~(\ref{Corr2}) in powers of~$g$, for which the following section provides evidence 
of a convergent tree-graph expansion. The second step consists of computing 
the free-boson correlators in~(\ref{Corr2}), which connects the leaves of the trees
in all possible ways, giving rise to the known UV divergencies and graph combinatorics. 
This approach thus amounts to a reorganization of the standard 
perturbative expansion: It removes all bosonic tadpoles and fermion loops,
effectively combining them into non-standard (hybrid) loops which 
in higher dimensions have the 
supersymmetry-induced cancellation of the leading UV~divergence 
already built in~\cite{FL,DL1,DL2,L1}.
Therefore, although integrating out the fermions does produce a highly nonlocal, and
thus seemingly more intricate, bosonic theory, thanks to the hidden supersymmetry
its correlation functions are not more complicated but potentially simpler than
those of the local formulation and offer new insights.

\subsection{Construction by dimensional reduction}

The goal of this section is the construction of $\cT_g$~\cite{Nic1,Nic2,FL,DL1,DL2,L1}
for the APD and SU($N$) supersymmetric matrix models~(\ref{LYM}) and (\ref{MMLag}).
This can be done either by repeating the construction procedure described
in~\cite{L1,Nic2} for this particular theory, or by dimensionally reducing the map
for ten-dimensional super Yang--Mills theory to one-dimensional matrix mechanics.
Let us first choose the second path.

Since we only have an on-shell formulation of supersymmetry in ten dimensions, 
we cannot employ the general scheme~\cite{MN,LR2} for arbitrary gauge fixing 
but have to stick to the Lorenz gauge, for which the map was presented 
on the gauge hypersurface in the critical spacetime dimensions $D=3, 4, 6$ and~$10$,
to $O(g^3)$ in~\cite{ALMNPP} and to $O(g^4)$ in~\cite{MN}.
In the dimensional reduction all quantities loose their coordinate dependence 
except for a dependence on time~$t$, and the $D$ components of the gauge potential
become $(D{-}1)$ dynamical matrices $X_a(t)$ and one non-dynamical matrix~$\om(t)$.
The Lorenz gauge reduces to $\del_t\om\equiv\dot\omega=0$, 
hence the matrix~$\omega$ is a constant on the gauge hypersurface. 
It will turn out that it is invariant under the map $\cT_g$.

Let us recall the salient facts of the construction, 
keeping $D$ arbitrary and denoting by~$r$ the dimension 
of the corresponding Majorana spinor representation.
The map $\cT_g$ is a nonlinear and nonlocal field transformation
\be
\cT_g: \bigl(X_a(t),\om\bigr)\ \mapsto\ \bigl(X'_a(t),\om'\bigr)\ .
\ee
It affords to express the quantum correlator $\langle F\rangle_g$ of an arbitrary
bosonic functional~$F$ at gauge coupling~$g$ in terms of a free correlator ($g{=}0$)
of the same functional, but with its arguments transformed by the inverse map,
\be
\Bigl\langle F[X,\om] \Bigr\rangle_g \ =\ 
\Bigl\langle F[\cT_g^{-1}X,\cT_g^{-1}\om] \Bigr\rangle_0\ .
\ee
An infinitesimal (in~$g$) version reads
\be \label{infflow}
\pa_g \Bigl\langle F[X,\om] \Bigr\rangle_g \ =\ 
\Bigl\langle \bigl( \pa_g + R_g[X,\om] \bigr) F[X,\om] \Bigr\rangle_g\ ,
\ee
where the ``coupling flow operator''~$R_g$ is a linear functional 
integro-differential operator
with a nonlinear and nonlocal dependence on $X$ and~$\om$.
As the construction is perturbative in the coupling~$g$,\footnote{
There exists, however, a universal nonperturbative formula for the map, see~\cite{LR1}.}
we expand (note the index shift)
\be\label{Rng}
R_g[X,\om] \ =\ \sum_{k=1}^\infty g^{k-1} \rR_k[X,\om] \ =\ 
\rR_1[X,\om] + g\,\rR_2[X,\om] + g^2 \rR_3[X,\om] + g^3 \rR_4[X,\om] + \ldots\ .
\ee
Integrating the infinitesimal flow equation (\ref{infflow}) yields $\cT_g^{-1}$ and finally
\be \label{TfromR}
\begin{aligned}
\cT_gX_a \ &=\ X_a\ -\ g\,\rR_1 X_a \ -\ \tfrac12g^2\bigl(\rR_2-\rR_1^2\bigr)X_a\ -\ 
\tfrac16g^3\bigl(2\rR_3-\rR_1\rR_2-2\rR_2\rR_1+\rR_1^3\bigr)X_a \\
&\quad -\ \tfrac{1}{24}\bigl(6\rR^4-2\rR_1\rR_3-3\rR_2^2+\rR_1^2\rR_2-6\rR_3\rR_1
+2\rR_1\rR_2\rR_1+3\rR_2\rR_1^2-\rR_1^4\bigr)X_a\ +\ \ldots
\end{aligned}
\ee
in terms of the flow operator's expansion coefficients.
We have displayed the result to $O(g^4)$ since we shall evaluate the map to this order,
and we omitted the analogous formula for~$\om$ because it reduces to $\cT_g\om=\om$.

In order to avoid cluttering the equations with indices, we mostly suppress spinor 
and color indices as well as time dependence and employ the DeWitt summation 
convention (suppressing also time integrals) in the remainder of this section.
We find it convenient to let the flow operator act (by functional differentiation) to the left.
It is then given by a variation $\overleftarrow{\frac{\d}{\d X^A_a(t)}}$ followed by a string
of matrices in color, spinor and coordinate space, such as
\be
(X_a{\times})^{AB}(t,t') = f^{AMB} X_a^M(t)\delta(t{-}t') \qquad\Rightarrow\qquad
(X_c{\times}X_d)^A(t) = f^{AMN} X_c^M(t) X_d^N(t)
\ee
and propagators $G$ and $S$ defined by
\be \label{props}
\bigl[D_t\,G\bigr]^{AB}_{\a\b}(t,t') = \d^{AB}\d_{\a\b}\d(t{-}t')\ ,\qquad
\bigl[(D_t+g\,\c^aX_a{\times})S\bigr]^{AB}_{\a\b}(t,t') = \d^{AB}\d_{\a\b}\d(t{-}t')
\ee
with $D_t\equiv D_0=\del_t+g\,\om{\times}$.
We note that $\om{\times}$ and $X_a{\times}$ are to be considered as matrices in color space.
The product of all these objects is to be executed in canonical fashion,
where we suppress obvious unit factors in the formul\ae. Observe also
that $G$ is {\em not\/} the ghost propagator whose defining equation
contains another derivative $\pa_t$. This is because in all relevant
expressions the ghost propagator appears with a derivative $\pa_t$.

A careful dimensional reduction of the coupling flow operator
eq.(1.19) of~\cite{ALMNPP} then yields~\footnote{
Note that we have split ${\cal R}=\del_g+R$.}
\be \label{Rright}
 \overleftarrow{R} \ =\ 
 -\tfrac{1}{r} \overleftarrow{\tfrac{\d}{\d X_a}}\,\tr\Bigl[
\bigl(\c_a-gX_a{\times}G\bigr)\,S\,
\bigl(\tfrac12\c^{cd}X_c{\times}X_d + \c^d\om{\times}X_d\bigr)\Bigr]
\ee
where the explicit trace refers to the spinor space,
and we have dropped a term proportional to~$\tfrac1g\dot{\om}$.
Here, the first round bracket arises from a non-abelian projector~\cite{ALMNPP}
which in the Lorenz gauge reads ($\mu=(t,a)$)
\be
P_\mu^{\ \nu} \ =\ \d_\mu^{\ \nu} - D_\mu (\del{\cdot}D)^{-1} \del^\nu
\qquad\stackrel{\textrm{reduction}}{\longrightarrow}\qquad
\d_\mu^{\ \nu} - D_\mu D_t^{-1}\del_t^{-1} \del^\nu
\ee
which obeys $\del^\mu P_\mu^{\ \nu} =0= P_\mu^{\ \nu} D_\nu$  and yields
\be \label{Pred}
P_t^{\ \nu}=0\ ,\qquad P_a^{\ b}=\d_a^{\ b} 
\qquad\textrm{and}\qquad
P_a^{\ t} = g\,X_a{\times}D_t^{-1}\ .
\ee
This shows that $R$ does not contain a variation $\frac{\d}{\d\om}$,
{\em cf.}~formula (1.19) in \cite{ALMNPP} (with $\mu = t$).
The second round bracket is just the decomposition of 
$\tfrac12 A_\rho{\times}A_\lambda$ in the reduction.

\subsection{Construction in the matrix model}

Alternatively, we may take the first path and construct the map $\cT_g$
directly for the matrix model, following the strategy of~\cite{L1,Nic2}.
To this end, we implement the Lorenz gauge constraint $\dot\om=0$
by adding to the matrix model Lagrangian~(\ref{MMLag})~\footnote{
Here and below, the $\cdot$ denotes a contraction in color space, i.e.~$P{\cdot}Q:=\d^{AB}P^A Q^B$.}
\be
\cL\ =\ \tfrac12(D_t X_a)^2\ -\ \tfrac14g^2(X_c{\times}X_d)^2 
\ -\ \tfrac{\mi}{2}\th\cdot(D_t+g\,\hat{X}{\times})\th
\ee
with $\hat{X}:=\c^aX_a$ a ``gauge-fixing term''
\be
\cL'\ =\ -\tfrac{1}{2\xi}\dot\om^2\ +\ \bar{C}\cdot\del_t D_t C
\ee
with a real parameter~$\xi$ and ghost matrices $C$ and~$\bar{C}$.
Taking the limit $\xi\ra 0$ puts the theory on the gauge hypersurface.

We aim to directly derive a coupling flow operator~$R$ as in~(\ref{infflow}) 
for the matrix model, which will govern the infinitesimal change in the coupling~$g$ 
for the quantum correlator of an arbitrary bosonic matrix functional~$F[X,\om]$.
Keeping in mind the $g$-dependence of the functional integral weight $\me^{\mi\int(\cL+\cL')}$,
we compute (suppressing the subscript in $\langle\cdots\rangle_g$)
\be \label{delF}
\begin{aligned}
\del_g\,\Bigl\langle \,F\,\Bigl\rangle \ &=\
\Bigl\langle \del_g F + F\,\del_g\int\!\mi(\cL{+}\cL') \Bigl\rangle \\
&=\ \Bigl\langle \del_g F + F\ \mi\int\!\Bigl[
D_t X_a\cdot\om{\times}X_a - \tfrac12g(X_c{\times}X_d)^2 
- \tfrac{\mi}{2}\,\th\cdot(\om+\hat{X}){\times}\th 
+ \bar{C}\cdot\del_t(\om{\times}C) \Bigr] \\
&=\ \Bigl\langle \del_g F + F\ \mi\Bigl[\delta_\a\Delta_\a 
+ \mi q\int\!\th{\cdot}(\om+\hat{X}){\times}\th
+ \int\!\bar{C}\cdot\del_t(\om{\times}C) \Bigr] \Bigl\rangle
\end{aligned}
\ee
where
\be \label{Delta}
\Delta_\a\ =\ -\tfrac{1}{r}\int\!\md t\;(\c^d\th)_\a{\cdot}\om{\times}X_d
\ +\ \tfrac{1}{2r}\int\!\md t\;(\c^{cd}\th)_\a{\cdot}X_c{\times}X_d\ .
\ee
With the supersymmetry transformations
\be
\d_\a\om = -\mi\,\th_\a\ ,\qquad \d_\a X_a = -\mi\,(\th\c_a)_\a\ ,\qquad
\d_\a \th_\b = -\c^d_{\a\b} D_t X_d -\tfrac{g}{2} \c^{cd}_{\a\b} X_c{\times} X_d
\ee
one confirms that indeed
\be
\d_\a\Delta_\a\ =\ \int\!\md t\;\Bigl[
D_t X_a\cdot\om{\times}X_a\ -\ \tfrac12g(X_c{\times}X_d)^2 
\ -\ \mi\,\tfrac{D-1}{r}\,\th\cdot(\om+\hat{X}){\times}\th \Bigr]\ .
\ee
Therefore, $\d_\a\Delta_\a$ in (\ref{delF}) reproduces $\del_g\int\!\cL$ 
but with a mismatch in the coefficient of the Majorana term, 
which thus still appears there but with a coefficient
\be
q\ =\ \tfrac{D-1}{r}-\tfrac12 \ =\ \tfrac1r \qquad\textrm{for}\quad D=3,4,6,10\ .
\ee
It is noteworthy that for a temporal gauge this mismatch is absent, 
\be
A_0 = 0 \qquad\Rightarrow\qquad \om=0 \quad\textrm{and}\quad D_t = \del_t
\qquad\Rightarrow\qquad \d_\a \bigl( \Delta_\a|_{\om=0} \bigr)\ =\ \del_g\int\cL\ ,
\ee
since effectively $D\to D{-}1$ and the ghosts decouple.

Next, we employ the broken supersymmetric Ward identity 
$\langle\d_\a Y\rangle=-\mi\langle(\d_\a\int\cL')\,Y\rangle$
together with
\be
\d_\a\,\int\!\cL' \ =\ -s\,\d_\a \Delta_{\textrm{gh}}
\qquad\textrm{for}\quad \Delta_{\textrm{gh}} \ =\ \int\!\bar{C}\,\dot{\om}
\ee
and the Slavnov variations
\be
s\,\om = D_t C\ ,\qquad s\,X_a = g\,X_a{\times}C\ ,\qquad s\,\th = g\,\th{\times}C\ ,\qquad
s\,C = -\tfrac{g}{2}C{\times}C\ ,\qquad s\,\bar{C} = \tfrac{1}{\xi}\dot\om
\ee
to rewrite
\be \label{dF}
\begin{aligned}
\del_g\,\Bigl\langle \,F\,\Bigl\rangle \ &=\
\Bigl\langle \del_g F + \mi\,\Delta_\a \d_\a F \Bigl\rangle \ +\
\Bigl\langle F\ \Bigl[ \Delta_\a\,s\,\d_\a \Delta_{\textrm{gh}}
- q\int\!\th{\cdot}(\om+\hat{X}){\times}\th
+ \mi\int\!\bar{C}\cdot\del_t(\om{\times}C) \Bigr] \Bigl\rangle\\
&=\ \Bigl\langle \del_g F + \mi\,\Delta_\a \d_\a F 
- \Delta_\a (\d_\a \Delta_{\textrm{gh}})\,s\,F \Bigl\rangle \\
&\ +\ \Bigl\langle F\ \Bigl[ \bigl(s\,\Delta_\a\bigr)\bigl(\d_\a\Delta_{\textrm{gh}}\bigr)
- q\int\!\th{\cdot}(\om+\hat{X}){\times}\th
- \mi\int\!\dot{\bar{C}}\cdot(\om{\times}C) \Bigr] \Bigl\rangle\ ,
\end{aligned}
\ee
where in the last step we used the BRST Ward identity $\langle s\,Y\rangle=0$.

For the flow equation~(\ref{infflow}) to hold, the last correlator has to vanish
for any bosonic functional~$F$. Writing out
\be
s\,\Delta_\a\ =\ \tfrac1r\int (\hat{X}{\times}\th)_\a\cdot\dot{C}
\qquad\textrm{and}\qquad
\d_\a \Delta_{\textrm{gh}} \ =\ -\mi\int\dot{\bar{C}}\cdot\th_\a
\ee
and performing the functional integrations over the fermions and the ghosts, 
this requirement becomes
\be \label{Z}
0\ \stackrel{!}{=}\ 
-\tfrac{\mi}{r}\int(\hat{X}_{\a\b}{\times}
\bcontraction{\th_b)\cdot}{\dot{C}}{\ \int}{\dot{\bar{C}}}
\bcontraction[2ex]{}{\th}{_\b)\cdot\dot{C}\ \int\dot{\bar{C}}\cdot}{\th}
\th_\b)\cdot\dot{C}\ \int\dot{\bar{C}}\cdot\th_\a
\ -\ q\int(\hat{X}_{\a\b}{\times}
\bcontraction{}{\th}{_\b)\cdot}{\th}
\th_\b)\cdot\th_\a
\ -\ q\int(\om{\times}
\bcontraction{}{\th}{_\a)\cdot}{\th}
\th_\a)\cdot\th_\a
\ +\ \mi\int\!
\bcontraction{(\om{\times}}{C}{)\cdot}{\dot{\bar{C}}}
(\om{\times}C)\cdot\dot{\bar{C}}\ ,
\ee
where the contractions stand for the fermionic and ghost propagators
\be \label{contractions}
\bcontraction{}{\th}{^B_\b(t)\ }{\th}
\th^B_\b(t)\ \th^A_\a(t')\ =\ -S^{BA}_{\b\a}(t,t') 
\qquad\textrm{and}\qquad
\bcontraction{}{C}{^B(t)\ }{\bar{C}}
C^B(t)\ \dot{\bar{C}}^A(t')\ =\ \mi\,G^{BA}(t,t')\ ,
\ee
respectively (note the time derivative on $\bar{C}^A$). With 
$\del_t G(t,t')^{BA}=\del_{t'}G^{AB}(t',t)=:\del G^{AB}(t',t)$ 
the condition~(\ref{Z}) reads
\be
0\ \stackrel{!}{=}\ 
-\tfrac1r\,\Tr\bigl[(\hat{X}{\times}S)\,\del G\bigr]
+q\,\Tr\bigl[ \hat{X}{\times}S \bigr]
+q\,\Tr\bigl[ \om{\times}S \bigr]
-\tfrac1r\,\Tr\bigl[\om{\times}G\bigr]\ ,
\ee
where the trace here refers to spin, color and time altogether.
Abbreviating the unit operator by the symbol $\unity$, and
inserting the useful identities
\be
\del G\ =\ \unity - g\,\om{\times}G
\qquad\textrm{and}\qquad
S\ =\ G - g\,G\,(\hat{X}{\times}S)\ ,
\ee
into the first and third term, respectively, we cancel the second and fourth terms
(provided $q=\tfrac1r$) and remain with
\be
0\ \stackrel{!}{=}\ 
\tfrac{1}{r}\,g\,\Tr\bigl[ (\hat{X}{\times}S)\,(\om{\times}G)\bigr]\ -\
q\,g\,\Tr\bigl[ (\om{\times}G)\,(\hat{X}{\times}S)\bigr]\ ,
\ee
which indeed holds in the critical dimensions.

We return to (\ref{dF}) and integrate out the fermions and ghosts
to read off the flow operator
\be
\begin{aligned}
R_g\ &=\ \mi
\bcontraction{}{\Delta}{_\a}{\d}
\Delta_\a \d_\a - 
\bcontraction{}{\Delta}{_\a(}{\d}
\Delta_\a (\d_\a 
\bcontraction{}{\Delta}{_{\textrm{gh}})\,}{s}
\Delta_{\textrm{gh}})\,s \\
&=\ 
\bcontraction{}{\Delta}{_\a \int\!}{\th}
\Delta_\a \int\!\th_\a\cdot\frac{\d}{\d\om}
\ +\ 
\bcontraction{}{\Delta}{_\a \int\!(}{\th}
\Delta_\a \int\!(\th\c_a)_\a\cdot\frac{\d}{\d X_a} \\
&\ -\ \mi
\bcontraction{}{\Delta}{_\a \int\!(}{\th}
\Delta_\a \int\!\th_\a\cdot
\bcontraction{}{\dot{\bar{C}}}{\int\!D_t}{C}
\dot{\bar{C}} \int\!D_t C\cdot\frac{\d}{\d\om}
\ -\ \mi
\bcontraction{}{\Delta}{_\a \int\!(}{\th}
\Delta_\a \int\!\th_\a\cdot
\bcontraction{}{\dot{\bar{C}}}{\int\!(g\,X_a{\times}}{C}
\dot{\bar{C}} \int\!(g\,X_a{\times}C)\cdot\frac{\d}{\d X_a}\ .
\end{aligned}
\ee
Since $D_t G=\unity$, the two variations w.r.t.~$\om$
(first and third terms) cancel, and we are left with
\be
R_g\ =\ 
\bcontraction{}{\Delta}{_\a \int}{\th}
\Delta_\a \int \th_\b \cdot \bigl[ (\c_a)_{\b\a}\,\unity
\ +\ g\,\d_{\b\a} G\times X_a \bigr] \cdot\frac{\d}{\d X_a}\ .
\ee
In the curly brackets we recognize the 
(dimensionally reduced) non-abelian projector~(\ref{Pred}).
Recalling $\Delta_\a$ from (\ref{Delta}), 
inserting the fermion propagator~(\ref{contractions})
and reversing the multiplication order, 
one again arrives at the flow operator presented in~(\ref{Rright}).

\subsection{The map to third and fourth order}

For the perturbative power series we need the expansion of the propagators,
\be
\begin{aligned}
G\ &=\ \ve - g\,\ve\,\om{\times}\ve + g^2\ve\,\om{\times}\ve\,\om{\times}\ve -
g^3\ve\,\om{\times}\ve\,\om{\times}\ve\,\om{\times}\ve \pm\ldots\ ,\\
S\ &=\ \ve - g\,\ve\,(\om+\hat{X}){\times}\ve + 
g^2\ve\,(\om+\hat{X}){\times}\ve\,(\om+\hat{X}){\times}\ve \mp\ldots\ ,
\end{aligned}
\ee
with the free fermion propagator (\ref{freeprop}).
Because spin traces vanish for an odd product of gamma matrices 
(for less than nine factors), the expansion coefficients $\rR_k$ displayed here carry
only even/odd powers of~$\om$ for $k$ being even/odd.
This parity extends to the map $\cT_g$ itself.
Carrying out the spin traces, one gets
\be
\begin{aligned}
\overleftarrow{\rR_1}\ &=\  -\overleftarrow{\tfrac{\d}{\d X_a}}\,\ve\,\om{\times}X_a \ ,\\
\overleftarrow{\rR_2}\ &=\  
\overleftarrow{\tfrac{\d}{\d X_a}}\,\ve\,\om{\times}\,\ve\,\om{\times}X_a +
\overleftarrow{\tfrac{\d}{\d X_a}}\,\ve\,X_b{\times}\,\ve\,X_b{\times}X_a\ , \\
\overleftarrow{\rR_3}\ &=\  
- \overleftarrow{\tfrac{\d}{\d X_a}}\,\ve\,\om{\times}\,\ve\,\om{\times}\,\ve\,\om{\times}X_a
- \overleftarrow{\tfrac{\d}{\d X_a}}\,\ve\,\om{\times}\,\ve\,X_b{\times}\,\ve\,X_b{\times}X_a
-\overleftarrow{\tfrac{\d}{\d X_a}}\,\ve\,X_b{\times}\,\ve\,\om{\times}\,\ve\,X_b{\times}X_a \\
&\ \quad
-\overleftarrow{\tfrac{\d}{\d X_a}}\,\ve\,X_a{\times}\,\ve\,X_b{\times}\,\ve\,\om{\times}X_b
+\overleftarrow{\tfrac{\d}{\d X_a}}\,\ve\,X_b{\times}\,\ve\,X_a{\times}\,\ve\,\om{\times}X_b
-\overleftarrow{\tfrac{\d}{\d X_a}}\,\ve\,X_b{\times}\,\ve\,X_b{\times}\,\ve\,\om{\times}X_a \\
&\ \quad
+\overleftarrow{\tfrac{\d}{\d X_a}}\,X_a{\times}\,\ve\,\ve\,X_b{\times}\,\ve\,\om{\times}X_b\ ,
\end{aligned}
\ee
and so on. Inserting all these into (\ref{TfromR}), performing the functional derivatives and
observing various cancellations, we arrive at
\be \label{TX}
\begin{aligned}
\cT_gX_a \ &=\ X_a\ +\ g\,\ve\,\om{\times}X_a \ -\ 
\tfrac12g^2 \ve\,X_b{\times}\,\ve\,X_b{\times}X_a \\[2pt]
&\ \ \ + \tfrac16g^3 \Bigl[
2\,\ve\,X_b{\times}\,\ve\,\om{\times}\,\ve\,X_b{\times}X_a +
2\,\ve\,X_a{\times}\,\ve\,X_b{\times}\,\ve\,\om{\times}X_b -
2\,X_a{\times}\,\ve\,\ve\,X_b{\times}\,\ve\,\om{\times}X_b \\
&\qquad\quad\ \ -
\ve\,X_b{\times}\,\ve\,X_a{\times}\,\ve\,\om{\times}X_b +
\ve\,X_b{\times}\,\ve\,X_b{\times}\,\ve\,\om{\times}X_a  +
\ve\,(\ve\,X_b{\times}X_a)\times(\ve\,\om{\times}X_b) \Bigr] \\
&\ \ \ +\ O(g^4)\ .
\end{aligned}
\ee
For the practitioner's convenience we spell this out with our shorthand notation fully 
expanded,
{\small
\be \label{TXlong}
\begin{aligned}
\cT_g X^A_a(t)\ &=\ X^A_a(t)\ +\ g\,f^{ABC}\!\int\!\!\md s\; \ve(t{-}s)\,\om^B X^C_a(s)\\
&\ -\ \tfrac12g^2 f^{ABC}f^{CDE}\!\int\!\!\md s\,\md u\; \ve(t{-}s)\,X^B_b(s)\,\ve(s{-}u)\,X^D_b X^E_a(u)\\
&\ +\ \tfrac16g^3 f^{ABC}f^{CDE}f^{EMN}\!\int\!\!\md s\,\md u\,\md v\; \ve(t{-}s)\,X^B_b(s)\,\ve(s{-}u)\,\Bigl[\\
&\quad 2\,\om^D(u)\,\ve(u{-}v)\,X^M_b X^N_a(v) - X^D_a(u)\,\ve(u{-}v)\,\om^M X^N_b(v) + X^D_b(u)\,\ve(u{-}v)\,\om^M X^N_a(v)
\Bigr]\\
&\ +\ \tfrac13g^3 f^{ABC}f^{CDE}f^{EMN}\!\int\!\!\md s\,\md u\,\md v\; 
\ve(t{-}s)\,\bigl[ X^A_a(s) - X^A_a(t)\bigr] \,\ve(s{-}u)\,X^D_b(u)\,\ve(u{-}v)\,\om^M X^N_b(v)\\
&\ +\ \tfrac16g^3 f^{ABC}f^{BDE}f^{CMN}\!\int\!\!\md s\,\md u\,\md v\; \ve(t{-}s)\,
\bigl[\ve(s{-}u)\,X^D_b X^E_a(u)\bigr]\bigl[\ve(s{-}v)\,\om^M X^N_b(v)\bigr] \\
&\ +\ \cO(g^4)\ .
\end{aligned}
\ee
}
As a check, beyond $\cO(g)$ all terms linear in~$X_a^A$ (and thus of maximal 
power in~$\om^A$) cancel out, a feature that can be proven to hold in general.
Also, in the temporal gauge $\om^A{=}0$ the map drastically simplifies and admits only 
even powers in~$g$ at least up to the order considered. Moreover,
to the order displayed here all terms share the ``linear tree" 
topology of the flow operator, except for the last term in $O(g^3)$, which is 
the first ``branched tree''. We have checked that this result is consistent with
the final result of~\cite{MN}. It is straightforward though tedious to extend the 
above computation to higher orders. Equivalently, the $\cO(g^4)$ result can be read off 
by dimensionally reducing the result of~\cite{MN}, but we refrain here from spelling 
out this formula with non-vanishing $\om$ because it is rather lengthy and not 
very illuminating. 

However, the result simplifies greatly in the temporal gauge $\om{=}0$. 
Furthermore, given our transcription rule~(\ref{fABC}), it is straightforward 
to write it down right away with the APD brackets~(\ref{APDbracket}).
Suppressing the common argument~$\bs$, we arrive at
\be\label{TXAPD}
\begin{aligned}
\cT_g X_a (t) \ &=\ X_a(t) \ -\  \tfrac12 g^2 \int\!\!\md s\,\md u\; \ve(t{-}s)\,\ve(s{-}u)\,
\Big\{ X_b(s) \,,\,\bigl\{ X_b(u) , X_a(u) \bigl\} \Big\} \\
&\quad +\ \tfrac18 g^4 \int\!\!\md s\,\md u\,\md v\,\md w\; \ve(t{-}s)\,\ve(s{-}u)\,\ve(u{-}v)\,\ve(v{-}w)\,\Biggl[ \\
&\qquad\qquad 6\ \biggl\{ X_b(s)\,,\,\Bigl\{ X_c(u)\,,\,\bigl\{ X_{[a}(v)\,,\,\{ X_b(w),X_{c]}(w)\}\bigr\}\Bigr\}\biggr\} \\
&\qquad\ \  +\ 2\ \biggl\{ X_b(s)\,,\,\Bigl\{ X_{[b}(u)\,,\,\bigl\{ X_{|c|}(v)\,,\,\{ X_{a]}(w),X_c(w)\}\bigr\}\Bigr\}\biggr\} \\
&\qquad\ \  +\ 2\ \biggl\{ X_a(s)-X_a(t)\,,\,\Bigl\{ X_b(u)\,,\,\bigl\{ X_c(v)\,,\,\{ X_b(w),X_c(w)\}\bigr\}\Bigr\}\biggr\}\,
\Biggr] \\
&\quad +\ \tfrac18 g^4 \int\!\!\md s\,\md u\,\md v\,\md w\; \ve(t{-}s)\,\ve(s{-}u)\,\ve(s{-}v)\,\ve(v{-}w)\,\times \\
&\qquad\qquad\ \ \biggl\{ \bigl\{ X_a(u) , X_b(u) \bigr\}\,,\,\Bigl\{ X_c(v)\,,\,\bigl\{ X_b(w) , X_c(w) \bigr\} \Bigr\}\biggr\}
\ \ +\ \cO(g^6)\ .
\end{aligned}
\ee
This expression is perfectly well-defined for well-behaved functions
$X_a(t,\bs)$, whence the $N\ra\infty$ limit of (\ref{TXlong}) is equally well-defined.
At higher orders we will encounter more nested APD brackets, but the expansion stays 
well-defined to arbitrary order. It is noteworthy that, while (\ref{TXlong}) contains both
even and odd powers in~$g$, the expansion (\ref{TXAPD}) with the temporal gauge $\om = 0$
contains only even powers in $g$. This can only change in higher orders (starting
with R$_7$ in (\ref{Rng}), to be completely precise) when we encounter 
$\c$-traces such as ${\rm tr}\, (\c^{a_1} \cdots \c^{a_9}) = 16\,\e^{a_1\cdots a_9}$. 

Finally, as an independent check, 
in the Appendix we also demonstrate that the ``free-action condition"
\be \label{freeaction}
\tfrac12(\del_t \cT_gX_a)^2 \ \stackrel{!}{=}\ \tfrac12(D_t X_a)^2 - \tfrac14g^2(X_b{\times}X_c)^2
+\ \textrm{total derivative}
\ee
as well as the ``determinant matching condition''~\footnote{
For the temporal gauge the last two terms on the r.h.s.~are replaced by
$\big[ \Tr\log D_t \,-\, \tfrac{D-1}{2}\Tr\log\del_t^2 \big]$, 
but the cancellations remain the same, of course.}
\be \label{detmatching}
\Tr\log\Bigl(\frac{\d \cT_gX}{\d X}\Bigr)
\ \stackrel{!}{=}\ 
\tfrac12\Tr\log\bigl(D_t+g\hat{X}{\times}\bigr)\ +\ \Tr\log\del_tD_t \,-\, \tfrac{D}{2}\Tr\log\del_t^2 
\ee
for the Jacobian, Matthews--Salam--Seiler~\cite{MS,S} and Faddeev--Popov~\cite{FP} determinants
are both fulfilled up to and including $O(g^3)$ by the maps~(\ref{TX}) and
(\ref{TXAPD}), provided
\be
D-2 \ \stackrel{!}{=}\ \frac{r}{2}\ ,
\ee
as happens to be the case for the critical dimensions $D=3,4,6$ and~$10$.
As already mentioned, the determinants are more subtle in the APD gauge theory directly:
like for the APD integral kernel (\ref{KkernelAPD}) we can in each APD bracket~(\ref{APDbracket})
separate the two $\bs$~arguments by inserting $\d$-functions together with integrals over $\bs$~variables.
When expanded, the Jacobian of the map $\cT_g$ then contains exactly the same divergent factor 
$\d(\bs,\bs)$ that we encountered in the expansion of $\D_{\rm MSS}$
and which can thus be dropped for the same reason.

The results of this section should be considered as a generalization of the polynomial map 
that obtains in supersymmetric quantum mechanics (see {\em e.g.}~\cite{Nic2}), where the 
perturbative expansion terminates after the first step and gives rise to a closed expression
(see also \cite{L2,AFF} for attempts to find polynomial maps in higher dimensions).
Such a feature cannot be expected for the  APD gauge theory or matrix model.
However, our expressions (\ref{TXlong}) and~(\ref{TXAPD}) are almost as good,
because they can be obtained from a universal formula for $\cT_g$ in terms of
a path-ordered exponential~\cite{LR1}. This formula furnishes an algorithmic
procedure to work out the expansion of $\cT_g$ systematically to any given order 
in~$g$, a calculation that can be automated and implemented on a  computer.
Again the result will be much simpler with the temporal gauge~$\om=0$.
On the technical side it is worth emphasizing that because of (\ref{axialLorenz})
the differences between the axial and the Lorenz gauge choices almost
disappear in one dimension, together with the considerable complications
accompanying gauge choices different from the Lorenz gauge in higher dimensions.

Let us, however, alert readers to a technical obstacle
that must be overcome before a perturbative evaluation of correlators
analogous to \cite{NP} can be set up.
In the limit of vanishing gauge coupling the measure in (\ref{Corr1}) becomes
{\em ultralocal\/} in $\bs$, with free propagators
\be
\big \langle X^a(t_1,\bs_1)\ X^b(t_2,\bs_2) \big\rangle_0
\ =\ \delta^{ab}\,C(t_1{-}t_2)\,\delta (\bs_1,\bs_2)
\ee
where $C(t)$ is the free scalar propagator in one time dimension. The reason is 
that, in the Lagrangian (\ref{LYM}), the spatial derivatives reside in the interaction 
term and are not part of the free measure. Consequently, one would have to sum 
an infinite number of terms to expose the full non-local structure of the theory.
As we remarked in the introduction and summary, this problem closely resembles 
the one of explaining the emergence of spatial structure from the ultralocal BKL limit
in quantum cosmology \cite{DHN}. In both cases, the `small tension limit' is
analogous to the one in the theory of elastic media when neighboring spatial
points become decoupled. Let us finally note that via the map $\cT_g$ the continuous 
spectrum of the interacting theory is almost self-evident from that of the free theory.

\section{The Jacobian has a non-zero radius of convergence}

One main difference between the present approach and more conventional
perturbative expansions of the path integral is that the series expansion 
for $\cT_g$ has better convergence properties (here we are not referring to
UV divergences, but to the non-summability of what would be the renormalized perturbation
expansion in higher dimensions).  That the convergence properties should be better 
was already anticipated in~\cite{Nic2} but never actually proven. Here we present further 
evidence for this conjecture by showing that with suitable technical assumptions
the Jacobian of the map admits a non-zero radius of convergence when expanded
around $g=0$ in the complex $g$~plane. This we can do by exploiting the equality
(\ref{detrelation}) of the Jacobian with (the product of) the fermionic determinants. 
Since we are actually only interested in the statement for the temporal gauge let us 
therefore set $\om = 0$ for which $\D_{\rm FP}$ is trivial, and consider the 
MSS determinant~(\ref{MSS}). To this aim we expand the logarithm of $\D_{\rm MSS}$ 
and make use of the triangle inequality,
\be\label{MSS1}
\Big| \log \det^{1/2}(\unity + gK) \Big| \,=\, \tfrac12\,
\Bigl|  \, \sum_{n=1}^\infty \tfrac{(-1)^{n-1}}{n}\, g^n\,\Tr\, K^n \Bigr|\ \leq \ 
\tfrac12 \, \sum_{n=1}^\infty \frac{|g|^n}{n} \, \bigl| \Tr\, K^n \bigr| \ ,
\ee
where the kernel $K$ is defined in (\ref{Kkernel}).
Let us have a look at the individual terms: we have
\be
\begin{aligned}
{\rm Tr} \, K^n \ =\ \int\!\md t_1 \cdots \int\!\md t_n \ &\ve(t_1{-}t_2) \cdots \ve(t_n{-}t_1) \
 {\rm tr} \bigl( \c^{a_1}\cdots \c^{a_n}\bigr) \, \times \\
 \times \  &{\rm tr} \bigl( T^{A_1} \cdots T^{A_n}\bigr) \
X_{a_1}^{A_1}(t_1) \cdots X_{a_n}^{A_n}(t_n)
\end{aligned}
\ee
where $T^A$ are the SU($N$) generators in the adjoint representation. We can now
derive an upper bound on the absolute value of this expression by 
using $\bigl|\ve(t)\bigr| \leq \frac12$, together with
\be
\bigl|\,\tr \bigl( \c^{a_1}\cdots \c^{a_n}\bigr) \bigr| \ \leq \ r  
\qquad \textrm{and} \qquad
\bigl|\,\tr \bigl( T^{A_1} \cdots T^{A_n}\bigr) \bigr| \ \leq \  c^n 
\ee
where $c\equiv c_N$ is an $N$-dependent positive constant. Furthermore
introducing the L$^1$-norm
\be
\big|\!\big| \bX \big|\!\big|_1 \ :=\ \sum_{a,A} \int\!\md t \ \bigl| X_a^A(t) \bigr|
\ee
we can majorize the individual terms to obtain
\be
 (\ref{MSS1}) \ \leq \ \frac{r}2 \, \sum_{n=1}^\infty \, \Bigl(\frac{c}2 \Bigr)^n 
 \frac{|g|^n}{n}\, \big|\!\big| \bX \big|\!\big|_1^n\ .
\ee
This series converges for $|g| < 2\,c^{-1} \big|\!\big| \bX \big|\!\big|_1^{-1}$.
Consequently if we constrain the functions $X_a^A(t)$ to belong 
each to the Lebesgue space L$^1(\Real)$, the series always has
an ($X$-dependent)  non-zero radius of convergence.

While the fact that the Jacobian has a non-zero radius of convergence as
a function of~$g$ does not imply that the map itself has this property, it strongly
constrains the series expansion for $\cT_g$, regardless of the precise 
form of the functions $X_a^A(t)$. 
The main reason that makes the argument work is that, unlike for higher-dimensional 
Yang--Mills theories, the supersymmetric matrix model has no UV divergences 
which would necessitate infinite subtractions (as in~\cite{S}). 
With appropriate UV and IR regularizations the above statements remain 
valid for supersymmetric Yang--Mills  theories in higher dimensions, at least 
with the axial gauge choice (for which, however, $\cT_g$ is considerably more
complicated than for the Lorenz gauge \cite{MN,LR2}). So in that case
{\em both\/} regulators are necessary  for the MSS~determinant to make
sense in a more rigorous context.

\newpage

\section{Outlook} 

We hope that the present investigations will open some new and so far unexplored
avenues for addressing several outstanding key problems of supermembrane and
matrix theory. Among the topics for future investigation we have already
highlighted two of these, namely the question of quantum target-space 
Lorentz invariance, and the problem of computing physically relevant 
correlations functions. Here our approach provides a perturbative expansion
scheme of a type that has not been available in the literature so far.
Finally we note that our methods may also turn out to be applicable
to matrix string theory \cite{M,DVV}, which corresponds to the reduction
of maximally extended super-Yang--Mills theory to two spacetime dimensions.

\bigskip\bigskip

\noindent
{\bf Acknowledgments.\ } 
We thank Hannes Malcha for help in matching (\ref{TXAPD}) with the dimensional 
reduction of the  $\cO(g^4)$ result of \cite{MN}, and Thibault Damour,
Daniele Dorigoni and Jan Plefka for discussions. We are also grateful 
to the referee for constructive criticism of an earlier version of this paper.


\newpage

\appendix
\section{Appendix: tests}

\subsection{Free action test}
Writing
\be
\cT_g X_a \ =\ X_a \ +\ g\,\rT_1X_a\ +\ g^2\rT_2X_a\ +\ g^3\rT_3X_a\ +\ O(g^4)
\ee
we read off from (\ref{TX}) the concrete expressions for~$\rT_k$.
The free-action condition~(\ref{freeaction}) then breaks up into
\be
\begin{aligned}
&\dot{X}_a\cdot(\rT_1X_a)^\bdot \ \stackrel{!}{=}\ \dot{X}_a\cdot(\om{\times}X_a)\ ,\\
&\tfrac12(\rT_1X_a)^\bdot \cdot (\rT_1X_a)^\bdot + \dot{X}_a \cdot (\rT_2X_a)^\bdot 
\ \stackrel{!}{=}\ \tfrac12(\om{\times}X_a)^2 - \tfrac14 (X_a{\times}X_b)^2\ ,\\
&(\rT_1X_a)^\bdot \cdot (\rT_2X_a)^\bdot + \dot{X}_a\cdot(\rT_3X_a)^\bdot
\ \stackrel{!}{=}\ 0\ ,
\end{aligned}
\ee
modulo total derivatives in~$t$.

The first condition is fulfilled since $(\rT_1X_a)^\bdot=\om{\times}X_a$.
This also matches the first terms on either side of the second condition.
Its remainder is also fulfilled because
\be
\begin{aligned}
\dot{X}_a \cdot (\rT_2X_a)^\bdot \ &=\ 
-\tfrac12 \dot{X}_a \cdot \bigl(X_b{\times}\,\ve\,X_b{\times}X_a\bigr)\ =\
-\tfrac12 \bigl(\dot{X}_a{\times}X_b\bigr)\cdot \ve\,\bigl(X_b{\times}X_a\bigr) \\
&=\ -\tfrac14 \bigl(X_a{\times}X_b\bigr)^\bdot \cdot\,\ve\,\bigl(X_b{\times}X_a\bigr)\ =\
-\tfrac14 \bigl(X_a{\times}X_b\bigr)^2\ +\ \del_t\bigl(\ldots\bigr)\ .
\end{aligned}
\ee
The third condition is more involved. The left-hand side reads
(suppressing total derivatives)
\be
\begin{aligned}
&-\ \tfrac12(\om{\times}X_a)\cdot X_b{\times}\,\ve\,(X_b{\times}X_a) \ +\
\tfrac16\dot{X}_a\cdot(\ve\,X_b{\times}X_a){\times}(\ve\,\om{\times}X_b) \\
&+\ \tfrac13\dot{X}_a\cdot \Bigl\{ 
X_b{\times}\,\ve\,\om{\times}\,\ve\,(X_b{\times}X_a) \ +\
X_a{\times}\,\ve\,X_b{\times}\,\ve\,(\om{\times}X_b) \ -\
\del_t\bigl\{( X_a{\times}\,\ve\,\ve\,X_b{\times}\,\ve\,(\om{\times}X_b) \bigr\} \\
&\qquad\quad -\ 
\tfrac12 X_b{\times}\,\ve\,X_a{\times}\,\ve\,(\om{\times}X_b) \ +\
\tfrac12 X_b{\times}\,\ve\,X_b{\times}\,\ve\,(\om{\times}X_a) \Bigr\} \\
=\ & -\ \tfrac12(\om{\times}X_a){\times} X_b\cdot\,\ve\,(X_b{\times}X_a) \ -\
\tfrac16 X_a\cdot\del_t\bigl\{(\ve\,X_b{\times}X_a){\times}(\ve\,\om{\times}X_b)\bigr\} \\
&+\ \tfrac13\dot{X}_a \times \Bigl\{
X_b\cdot\,\ve\,\om{\times}\,\ve\,(X_b{\times}X_a) \ -\
\dot{X}_a\cdot\,\ve\,\ve\,X_b{\times}\,\ve\,(\om{\times}X_b) \ -\
X_b\cdot\,\ve\,X_{[a}{\times}\,\ve\,(\om{\times}X_{b]}) \Bigr\} \\
=\ & -\ \tfrac12(\om{\times}X_a){\times} X_b\cdot\,\ve\,(X_b{\times}X_a) \ -\
\tfrac16 X_a\cdot(X_b{\times}X_a){\times}\,\ve\,(\om{\times}X_b) \ +\
\tfrac16 X_a\cdot(\om{\times}X_b){\times}\,\ve\,(X_b{\times}X_a) \\
&+\ \tfrac16 (X_a{\times}X_b)^\bdot\cdot\,\ve\,\om{\times}\,\ve\,(X_b{\times}X_a) \ -\
\tfrac16 (X_a{\times}X_b)^\bdot\cdot\,\ve\,X_a{\times}\,\ve\,(\om{\times}X_b) \\
=\ & -\  \tfrac12(\om{\times}X_a){\times} X_b\cdot\,\ve\,(X_b{\times}X_a) \ -\
\tfrac16 X_a{\times}(X_b{\times}X_a)\cdot\,\ve\,(\om{\times}X_b) \ +\
\tfrac16 X_a{\times}(\om{\times}X_b)\cdot\,\ve\,(X_b{\times}X_a) \\
&-\ \tfrac16 (X_a{\times}X_b)\cdot\om{\times}\,\ve\,(X_b{\times}X_a) \ +\
\tfrac16 (X_a{\times}X_b)\cdot X_a{\times}\,\ve\,(\om{\times}X_b) \\
=\ & -\ \tfrac16 \Bigl\{
3\,(\om{\times}X_a){\times} X_b\ -\ X_a{\times}(\om{\times}X_b)\ +\ (X_a{\times}X_b){\times}\om
\Bigr\}\cdot\,\ve\,(X_b{\times}X_a)\\
=\ & -\ \tfrac16 \Bigl\{
3\,(\om{\times}X_{[a}){\times}X_{b]}\ +\ (\om{\times}X_{[b}){\times}X_{a]}\ +\ (X_a{\times}X_b){\times}\om
\Bigr\}\cdot\,\ve\,(X_b{\times}X_a)\\
=\ & -\ \tfrac16 \Bigl\{
(\om{\times}X_a){\times}X_b\ +\ (X_b{\times}\om){\times}X_a\ +\ (X_a{\times}X_b){\times}\om 
\Bigr\}\cdot\,\ve\,(X_b{\times}X_a) \ =\ 0\ ,
\end{aligned} 
\ee
where in the last line the Jacobi identity was applied.
Several times we employed partial integration and $\del_t\,\ve=\unity$ as well as
$A\cdot(B{\times}C)=(A{\times}B)\cdot C$ 
and the complete antisymmetry of the structure constants, $A{\times}B=-B{\times}A$.
Furthermore, for the first equality we cancelled part of the $\del_t$ term with the term preceding it,
for the second equality we dropped a term $\sim\dot{X}_a{\times}\dot{X}_a=0$,
for the fourth equality the second and fifth terms cancelled, 
and for the fifth equality the index antisymmetry in the final factor $X_b{\times}X_a$ was used.
We note that the value~$D$ of the spacetime dimension played no role here.

\subsection{Determinant matching test}
Since the determinants match in the free theory, it suffices to bring their logarithms to a form
\be \label{lndet}
\log\det\Delta(g)\ =\ \log\det\Delta(0)\ +\ \Tr\log\bigl(\unity+M(g)\bigr)
\ =\ \textrm{const} + \Tr M - \tfrac12\Tr M^2 + \tfrac13\Tr M^3 +O(g^4)
\ee
since $M(g)$ is of order~$g$, and to compare the expressions in the orders 
$g$, $g^2$ and $g^3$ of the perturbative expansion.
The $\Tr$ symbol refers to a trace in position, color and spinor space,
while below we reserve the $\tr$ symbol for the trace in position and color space only,
after having explicitly performed the gamma traces.

For the Faddeev--Popov determinant we have
\be
\Delta = \del_t D_t = \del_t(\del_t+g\,\om{\times}) \qquad\Rightarrow\qquad 
M = g\,\del_t^{-1}\,\om{\times} = g\,\ve\,\om{\times}
\ee
which, since there are no spin degreees of freedom, leads to
\be
\tr\log\bigl(\unity+M(g)\bigr) \ =\
g\;\tr\,\bigl(\om{\times}\,\ve\bigr)\ -\ 
\tfrac12 g^2\,\tr\,\bigl(\om{\times}\,\ve\,\om{\times}\,\ve\bigr)\ +\ 
\tfrac13 g^3\,\tr\,\bigl(\om{\times}\,\ve\,\om{\times}\,\ve\,\om{\times}\,\ve\bigr)\ +\
O(g^4)\ .
\ee
The Matthews--Salam--Seiler determinant produces
\be
\Delta = D_t + g\,\hat{X}{\times} = \del_t + g\,(\om+\hat{X}){\times} \qquad\Rightarrow\qquad
M = g\,\ve\,(\om+\hat{X}){\times}
\ee
which, with a factor of~$\tfrac12$ from the Majorana property, yields
\be
\begin{aligned}
\tfrac12\Tr\log\bigl(\unity+M(g)\bigr) \ &=\
\tfrac{r}{2}g\;\tr\,\bigl(\om{\times}\,\ve\bigr)\ -\ 
\tfrac{r}{4}g^2\,\tr\,\bigl(\om{\times}\,\ve\,\om{\times}\,\ve\ +\ X_a{\times}\,\ve\,X_a{\times}\,\ve \bigr) \\
&\ +\ \tfrac{r}{6}g^3\,\tr\,\bigl(\om{\times}\,\ve\,\om{\times}\,\ve\,\om{\times}\,\ve\ +\ 
3\, X_a{\times}\,\ve\,X_a{\times}\,\ve\,\om{\times}\,\ve \bigr)\ +\
O(g^4)\ ,
\end{aligned}
\ee
where only even powers of~$\hat{X}$ survived the spin trace, which produces a factor~$r$ for the
dimensionality of the spinor representation.
Each of the trace terms can be represented by a loop diagram, with bosonic propagators~$\ve$ 
and external ``legs'' $\om$ or $X_a$. Because $\ve(t{-}t)=\ve(0)=0$, single-leg loops vanish,
and we only have to consider the orders $g^2$ and~$g^3$ in the matching.

Finally, considering the Jacobian of $\cT_g$,
we must in each tree of the expression~(\ref{TX}) ``differentiate away'' 
one ``leaf''~$X$ in all possible ways.
This results in an expression of the form
\be
\frac{\d \cT_gX^A_a(t)}{\d X^B_b(t')} \ =\ \d^{AB}\d_{ab}\d(t{-}t')\ +\ 
\bigl( g\,M_1 + g^2 M_2 + g^3 M_3 +O(g^4)\bigr)^{AB}_{ab}(t,t')\ ,
\ee
which can be viewed as a string starting from the tree root and ending at the cut leaf location,
possibly with branches attached to it.
Inserting this expansion into (\ref{lndet}) we find
\be
\log\det\Bigl(\frac{\d \cT_gX}{\d X}\Bigr)\ =\ \textrm{const}\ +\ 
g\,\Tr M_1\ +\ g^2\bigl(\Tr M_2 - \tfrac12\Tr M_1^2\bigr)\ +\ 
g^3\bigl(\Tr M_3 - \Tr M_1M_2 + \tfrac13\Tr M_1^3 \bigr)\ .
\ee
Under each trace we glue together the strings in the product
and then short-circuit the total string by identifying the end points and summing over 
the corresponding indices (including integration over time).
As a result we collect
\be
\begin{aligned}
\Tr M_2\ &=\ -\tfrac12(D{-}2)\,\tr\,\bigl(X_a{\times}\,\ve\,X_a{\times}\,\ve\bigr)\ , \\
- \tfrac12\Tr M_1^2\ &=\ -\tfrac12(D{-}1)\,\tr\,\bigl(\om{\times}\,\ve\,\om{\times}\,\ve\bigr)
\end{aligned}
\ee
and, after several cancellations,
\be
\begin{aligned}
\Tr M_3\ &=\ 
(\tfrac{D}{2}{-}\tfrac23)\,\tr\,\bigl(X_a{\times}\,\ve\,X_a{\times}\,\ve\,\om{\times}\,\ve\bigr)\ +\
\tfrac13\,\tr\,(\ve\,X_a{\times}\,\ve){\times}\,\ve\,\om{\times}X_a \\
&\quad -\ \tfrac13\,\tr\,\bigl(\ve\,\,\ve\,X_a{\times}\bigr){\times}\,\ve\,\om{\times}X_a\ -\
\tfrac13\,\tr\,\bigl(X_a{\times}\,\ve\,X_a{\times}\,\om{\times}\,\ve\,\,\ve\bigr) \\
&=\ (\tfrac{D}{2}{-}1)\,\tr\,\bigl(X_a{\times}\,\ve\,X_a{\times}\,\ve\,\om{\times}\,\ve\bigr)\ +\
\tfrac13\,\tr\,(\ve\,X_a{\times}\,\ve){\times}\,\ve\,\om{\times}X_a\ ,\\
- \Tr M_1M_2\ &=\ (\tfrac{D}{2}{-}1)\,\tr\,\bigl(X_a{\times}\,\ve\,X_a{\times}\,\ve\,\om{\times}\,\ve\bigr)\ ,\\
\tfrac13\Tr M_1^3\ &=\ \tfrac13(D{-}1)\,\tr\,\bigl(\om{\times}\,\ve\,\om{\times}\,\ve\,\om{\times}\,\ve\bigr)\ .
\end{aligned}
\ee
In the four contributions to $\Tr M_3$, the fourth term is of the same form as the first one
because $\om$ being constant can be moved  past $\ve$. 
The other two contributions are loops with a branch attached.
The third term vanishes because the trace is proportional to $\del_t^{-2}(0)$
which gets regularized to zero. 
Finally, the second term is of the form
\be
f(t',t'') \int\!\md t\ \ve(t{-}t')\,\ve(t'{-}t)\,\ve(t{-}t'') \ =\ -\tfrac14\,f(t',t'')\int\!\md t\ \ve(t{-}t'') \ =\ 0\ .
\ee

Collecting all remaining contributions, we end up with two 2-leg loops at $O(g^2)$
and two 3-leg loops at $O(g^3)$:
\begin{center}
\begin{tabular}{|l|c|c|c|}
\hline
expression & FP & MSS & Jac \\
\hline $\vphantom{\Big|}
g^2\,\tr\,\bigl(\om{\times}\,\ve\,\om{\times}\,\ve\bigr)$ 
& $-\tfrac12$ & $-\tfrac{r}{4}$ & $\tfrac12(1{-}D)$ \\[4pt]
$g^2\,\tr\,\bigl(X_a{\times}\,\ve\,X_a{\times}\,\ve\bigr)$ 
& $0$ & $-\tfrac{r}{4}$ & $\tfrac12(2{-}D)$ \\[4pt]
\hline $\vphantom{\Big|}
g^3\,\tr\,\bigl(\om{\times}\,\ve\,\om{\times}\,\ve\,\om{\times}\,\ve\bigr)$ 
& $\tfrac13$ & $\tfrac{r}{6}$ & $\tfrac13(D{-}1)$ \\[4pt]
$g^3\,\tr\,\bigl(X_a{\times}\,\ve\,X_a{\times}\,\ve\,\om{\times}\,\ve\bigr)$ 
& $0$ & $3\,\tfrac{r}{6}$ & $D{-}2$ \\[4pt]
\hline
\end{tabular}
\end{center}
Here, ``FP'', ``MSS'' and ``Jac'' denote the weight of the individual expressions contributing
to the logarithm of the Faddeev--Popov, Matthews--Salam--Seiler and Jacobian determinant, 
respectively. Fortunately, the sum of the FP and MSS columns agrees with the Jac column
provided that again $D{-}2=\tfrac{r}{2}$, singling out the critical dimensions once more. 
This provides a nontrivial check on the expression~(\ref{TXlong}) of the Nicolai map,
which formally is guaranteed to work out by the construction scheme. 
Finally, we remark that the matching also works in the temporal gauge,
since the Faddeev--Popov determinant becomes trivial but the first and third expression
in the table vanish for~$\om=0$ anyway.

\end{document}